\documentclass[a4paper,showpacs,aps,draft,floatfix] {revtex4}
\input epsf

\newcounter{subequation}[equation]
\makeatletter
\fussy
\def \L {\Lambda }
\def \l {\lambda } 
 
\def \t {\theta }
\def \vt {\vartheta }

\def \a {\alpha }

\def \d {\delta }
\def \D {\Delta }

\def \g {\gamma }
\def \G {\Gamma }
\def \O {\Omega }

\def \b {\beta }

\def \s {\sigma }
\def \e {\epsilon }

\def \ud { {1 \over 2} }

\def \cddd { {\cal D } }
\def \cala { {\cal A } }
\def \calc { {\cal C } }
\def \calp { {\cal P } }

\def \calf { {\cal F } }

\def \caln { {\cal N } }
\def \call { {\cal L } }

\def \calT { {\cal T } }

\def \calr { {\cal R } }
\def \cals { {\cal S } }

\def \cals { {\cal S } }
\def \calv { {\cal V } }

\def \Eslash {E \kern-.5em\slash}
\def \epslash {\epsilon \kern-.5em\slash}
\def \pslash {p \kern-.5em\slash}
\def \kslash {k \kern-.5em\slash}
\def \Dslash {D \kern-.5em\slash}
\def \hslash {h \kern-.5em\slash}
\def \dslash {\partial \kern-.5em\slash}
\def \vslash {v \kern-.5em\slash}

\def \bfA { {\bf A} }

\def \bfS { {\bf S} }

\def\NPB#1#2#3 {{\rm Nucl.~Phys.}  {\bf{B#1}}, {#3} (#2)}
\def\NPA#1#2#3 {{\rm Nucl.~Phys.}  {\bf{A#1}}, {#3} (#2)}
\def\PLB#1#2#3 {{\rm Phys.~Lett.}  {\bf{B#1}}, {#3} (#2)}
\def\PR#1#2#3 {{\rm Phys.~Rep.}  {\bf#1}, {#3} (#2)} 
\def\PRD#1#2#3 {{\rm Phys.~Rev.}  {\bf{D#1}}, {#3} (#2)} 
\def\PRL#1#2#3 {{\rm Phys.~Rev.~Lett.}  {\bf{#1}}, {#3} (#2)} 
\def\ZPC#1#2#3 {{\rm Z.~Phys.}  {\bf C#1}, {#3} (#2)} 
\def\JHEP#1#2#3 {{\rm JHEP} {\bf C#1}, {#3}  (#2)}     
\def\IJMP#1#2#3 {{\rm Int. J. Mod. Phys.}  {\bf A#1}, {#3} (#2)} 
\def\JMP#1#2#3 {{\rm J. Math. Phys.}  {\bf #1}, {#3} (#2)}
\def\PTP#1#2#3 {{\rm Prog. Theor. Phys.}  {\bf{#1}}, {#3} (#2)}
\def\FP#1#2#3  {{\rm  Fortsch. Phys.}  {\bf{#1}}, {#3} (#2)}
\def\CMP#1#2#3  {{\rm Comm. Math. Phys.}  {\bf{#1}}, {#3} (#2)}

\makeatletter

\def\thesubequation{\theequation\@alph\c@subequation}
\def\@subeqnnum{{\rm (\thesubequation)}}
\def\slabel#1{\@bsphack\if@filesw {\let\thepage\relax
   \xdef\@gtempa{\write\@auxout{\string
      \newlabel{#1}{{\thesubequation}{\thepage}}}}}\@gtempa
   \if@nobreak \ifvmode\nobreak\fi\fi\fi\@esphack}
\def\subeqnarray{\stepcounter{equation}
\let\@currentlabel=\theequation\global\c@subequation\@ne
\global\@eqnswtrue
\global\@eqcnt\z@\tabskip\@centering\let\\=\@subeqncr
$$\halign to \displaywidth\bgroup\@eqnsel\hskip\@centering
  $\displaystyle\tabskip\z@{##}$&\global\@eqcnt\@ne
  \hskip 2\arraycolsep \hfil${##}$\hfil
  &\global\@eqcnt\tw@ \hskip 2\arraycolsep
  $\displaystyle\tabskip\z@{##}$\hfil
   \tabskip\@centering&\llap{##}\tabskip\z@\cr}
\def\endsubeqnarray{\@@subeqncr\egroup
                     $$\global\@ignoretrue}
\def\@subeqncr{{\ifnum0=`}\fi\@ifstar{\global\@eqpen\@M
    \@ysubeqncr}{\global\@eqpen\interdisplaylinepenalty \@ysubeqncr}}
\def\@ysubeqncr{\@ifnextchar [{\@xsubeqncr}{\@xsubeqncr[\z@]}}
\def\@xsubeqncr[#1]{\ifnum0=`{\fi}\@@subeqncr
   \noalign{\penalty\@eqpen\vskip\jot\vskip #1\relax}}
\def\@@subeqncr{\let\@tempa\relax
    \ifcase\@eqcnt \def\@tempa{& & &}\or \def\@tempa{& &}
      \else \def\@tempa{&}\fi
     \@tempa \if@eqnsw\@subeqnnum\refstepcounter{subequation}\fi
     \global\@eqnswtrue\global\@eqcnt\z@\cr}
\let\@ssubeqncr=\@subeqncr
\@namedef{subeqnarray*}{\def\@subeqncr{\nonumber\@ssubeqncr}\subeqnarray}
\@namedef{endsubeqnarray*}{\global\advance\c@equation\m@ne%
                           \nonumber\endsubeqnarray}

\@addtoreset{equation}{section}
\renewcommand{\theequation}{\thesection.\arabic{equation}}

\newcommand{\be}{\begin{equation}} 
\newcommand{\ee}{\end{equation}} 
\newcommand{\ba}{\begin{array}}
\newcommand{\ea}{\end{array}}
\newcommand{\bea}{\begin{eqnarray}} 
\newcommand{\eea}{\end{eqnarray}} 
\newcommand{\bsea}{\begin{subeqnarray}} 
\newcommand{\esea}{\end{subeqnarray}}

\begin{document}
\rightline{IPhT-Saclay Preprint t08/145} 
\title{\bf Contact interactions in low scale string models with
intersecting $D6$-branes} \author{M. Chemtob}
\email{marc.chemtob@cea.fr} \affiliation{ CEA, DSM, Institut de
Physique Th\'eorique, IPhT, CNRS, MPPU, URA2306, Saclay, F-91191
Gif-sur-Yvette, FRANCE} \thanks {\it Supported by the Laboratoire de
la Direction des Sciences de la Mati\`ere du Commissariat \`a
l'Energie Atomique } \date{\today}


\begin{abstract} 
We evaluate the tree level four fermion string amplitudes in the TeV
string mass scale models with intersecting $D6$-branes.  The
coefficient functions of contact interactions subsuming the
contributions of string Regge resonance and winding mode excitations
are obtained by subtracting out the contributions from the string
massless and massive momentum modes.  Numerical applications are
developed for the Standard Model like solution of Cremades, Ibanez,
and Marchesano for a toroidal orientifold with four intersecting
$D6$-brane stacks.  The chirality conserving contact interactions of
the quarks and leptons are considered in applications to high energy
collider and flavor changing neutral current phenomenology.  The two
main free parameters consist of the string and compactification mass
scales, $ m_s$ and $ M_c$.  Useful constraints on these parameters are
derived from predictions for the Bhabha scattering differential cross
section and for the observables associated to the mass shifts of the
neutral meson systems $ K-\bar K ,\ B-\bar B,\ D-\bar D$ and the
lepton number violating three-body leptonic decays of the charged
leptons $\mu $ and $ \tau $.
\end{abstract} \maketitle

\section{INTRODUCTION}
\label{sect1}

The consideration of Dirichlet branes has led in recent years to
remarkable advances in the particle physics model building. This has
made possible the construction of wide classes of Standard Model
realizations for type $II$ superstring theories using branes which
extend along the flat spatial dimensions of $M_4$ and wrap around
cycles of the internal space manifold.  The existing two approaches
employing configurations of multiple type $II b $ branes located near
orbifold singularities~\cite{gimpol,gijohn,singmod} and type $II a $
branes intersecting at angles~\cite{berk96}, which we designate
henceforth for lack of better names as setups of branes within branes
and intersecting branes, respectively, are well-documented by now,
thanks to the reviews in~\cite{revbib}
and~\cite{blumshiu05,blumger07,marche07}.  The two most characteristic
features of these string constructions reside in the wider freedom in
choosing the string theory mass scale and in the occurrence of
localized chiral fermions in the string spectrum.

Because they are amenable to experimental tests based on data from
high-energy colliders and limits for rare processes, the TeV string
mass scale models~\cite{antb90,lykken96} are clearly those with the
greatest impact on phenomenology. In developing the string theory
machinery for the physics from extra dimensions, one is especially
encouraged by the applications using orbifold field theories in higher
dimensional spacetimes with matter fermions which move in the bulk or
are localized inside thin~\cite{add99,randrum99} or
thick~\cite{arkan99,arkan00,mira00,georhailu01} domain wall branes.
The theoretical and experimental aspects of collider studies of
physics from extra dimensions are reviewed
in~\cite{han99,giudice99,hewett02,cheung03} and~\cite{lands04}.  The
string theory framework embodies a higher degree of consistency in
comparison to the field theory framework, and is also more economical
thanks to a small parameter space restricted to the fundamental
tension and coupling constant string parameters, $ m ^2_s =1/\a '$ and
$g_s$, along with adjustable parameters associated with the
compactification and infrared cutoff mass scales.  Several examples
that qualify as TeV scale string models have been constructed within
the branes within branes~\cite{kakutev98} and intersecting
branes~\cite{isbtev} approaches.

One important motivation for the interest in TeV scale string models
is to gain insight on the hierarchy between the contributions from the
various string excitations.  In particular, one wishes to understand
how the exchange of string Regge and winding modes compares with that
of the gravitational Kaluza-Klein (KK) modes. This has a practical
importance since, unlike the contributions the latter ones can be well
described, in principle, within the familiar field theory framework.

The early string inspired studies were focused on models using single
brane configurations~\cite{dudas0,cpp00}. These have the
characteristic property that the open string Regge resonance modes
contribute at the tree (disk surface) level while the closed string
modes contribute at the one-loop (cylinder surface) level of the
string perturbation theory on the world sheet.  (That the exchange of
closed string modes can be viewed as a loop effect follows from the
world sheet duality linking the ultraviolet and infrared regimes.)
The strong constraints from the $\caln =4$ supersymmetry preserved by
the extremal branes of type II supergravity which saturate the BPS
(Bogomolny-Prasad-Sommerfeld) bound, restrict the low-energy tree
level contributions in single brane models to local operators of
dimension $ \cddd \geq 8 $.  Using the orbifolding mechanism, say, by
placing the $D$-brane at an orbifold fixed point in order to reduce
the number of preserved 4-d supercharges and produce a chiral open
string spectrum, is indeed helpful towards obtaining semirealistic
models but does not significantly modify the structure of string
amplitudes. This is often referred as the inheritance property of
orbifolds.  Another characteristic of the single $D$-brane models is
the insensitivity of predictions with respect to the structure of the
internal space manifold.  Here again the orbifold constraints on the
Chan-Paton (CP) gauge factors greatly improve the description by
ensuring that massless poles in channels with exotic quantum numbers
cancel out.  Useful applications to neutrino-nucleon elastic
scattering at ultra-high energies~\cite{cornet01,friess02} and to the
two-body reactions at high-energy
colliders~\cite{burik03,burik04,burik06} have been developed along
these lines in terms of single $D$-brane models where $m_s$ and the CP
factors are treated as free parameters. Also, building up on the
initial studies of the single photon+jet signal in the high energy
hadronic collider reaction, $p + p\to \g + j$, recent works discuss
the dijet signal~\cite{ancho08} in the reaction, $ p + p \to j + j $,
along with signals from the various processes at the LHC
model~\cite{lust08} which could usefully test the single $D$-brane
models of TeV string scale..

The multiple brane setups bring a crucial novel feature in this
discussion through the presence of localized massless fermions whose
contact interactions are not restricted to dimensions $ \cddd \geq 8
$.  This point was first recognized by Antoniadis et
al.,~\cite{antbengier01} in the context of TeV scale string models
with $Dp/D(p+4)$-branes.  Finite tree level contributions were indeed
found for the dimension $ \cddd =6 $ local operators coupling four
fermions of which at least a single pair belongs to the non-diagonal
open string sectors $(p, p+4) + ( p+4,p) $.  That the open string
fermions localized at the intersection of $ D$-branes behave in much
the same way as the twisted modes of closed
strings~\cite{bersh87,dixon87} was realized by several authors in the
context of multiple brane~\cite{hashi96,gava97,fro099} and
intersecting
brane~\cite{Cvetic:2003ch,owen02,owen03,owenp03,KW03,chem06} models.
The operator algebra approach for the superconformal field theory on
the world sheet can be used to calculate the tree level string
amplitudes.  We also note as a side remark that the tools developed in
recent years for the calculation of scattering amplitudes in gauge
theories~\cite{cachazo05} and string theories~\cite{stietaylor07}
should encourage pursuing applications for general $n$-point
amplitudes at higher loop orders.

So far, the collider studies of TeV string scale models have been
mostly focused on setups with single branes~\cite{dudas0,cpp00} and
branes within branes~\cite{antbengier01,antorev01,shiusch99}, as
already said above.  No comparable applications exist for the
intersecting branes models.  Regarding the flavor physics, however,
several studies have been devoted to both the branes within
branes~\cite{kakutev98} and the intersecting
branes~\cite{abel03,abeleb03} models.  In parallel, a wide interest
was aroused by the field theory models in flat extra dimensions with
thin branes~\cite{delgado99,rizzo01} and thick
branes~\cite{splitf03,masiero01,baren01,lill03,lill203} or in warped
spacetimes~\cite{moreau06,csaki08}.  The reader should be warned that
the quoted references represent a tiny fraction of the literature on
this subject.

In the present work we wish to pursue the discussion of the tree level
contact interactions for four localized fermions in intersecting brane
models with the view to confront the predictions against experimental
data for colliders and flavor changing neutral current processes.
Although the assumption of a low string mass scale is naturally paired
with that of large extra space dimensions, the requirement that the
string theory remains weakly coupled turns out to restrict the ratio
of string to compactification scales, $ m_s r= m_s / M_c$, to a
relatively narrow interval of $ O(1)$.  This circumstance has
motivated us in taking the contributions from world sheet instantons
into account.  We evaluate the tree level string amplitudes by
integrating the vacuum world sheet correlators over the moduli space
of the disk surface with two pairs of massless fermion vertex
operators inserted on the boundary.  Since the tools for calculating
open string amplitudes in intersecting brane
models~\cite{hashi96,gava97,fro099,Cvetic:2003ch,owen03,KW03} are well
documented by now, we shall present very briefly the main formulas
before proceeding to our main goal. The discussion will rely heavily
on our previous work~\cite{chem06}.

We develop concrete calculations for the Standard Model solution
obtained by Cremades et al.,~\cite{CIM,CIM2} for a toroidal
orientifold with four $D6$-branes, using specifically the related
family of solutions presented by Kokorelis~\cite{kokosusy02}.  This is
conceived as a local model (or premodel) described by a classical
configuration of intersecting $D6$-branes decoupled from the
gravitational interactions and the geometric moduli fields.  It is
encouraging that other searches of solutions for orientifolds with
intersecting branes also select families of small size, as illustrated
by the study focused on the $ Z_2 \times Z_2$ orbifold
models~\cite{Cvetic:2004ui} realizing the supersymmetric Pati-Salam
model with a hidden sector, and the statistical studies of the open
string landscape of vacua for the $ Z_2 \times
Z_2$~\cite{gmeiner05,douglor06} and $ Z_{6-II} $~\cite{gmeiner08}
orbifolds realizing the minimal supersymmetric standard model and the
Pati-Salam model with hidden sectors.

The proper identification of physics from extra dimensions presupposes
that one can combine the new physics contributions with those from the
Standard Model interactions in a consistent way.  This condition is
especially critical for the string theory applications where a
satisfactory implementation of the electroweak symmetry breaking is
not yet available.  Rather than pursuing a full-fledged calculation,
we shall adopt here a phenomenologically minded approach, similar to
that used in~\cite{antbengier01}.  This consists in separating out by
hand in the low-energy expansion of string amplitudes the
contributions from the string massless and momentum modes so as to
access the contact interactions which subsume the contributions from
the string Regge and winding excitations.

The exchange of massive modes from extra dimensions can also induce
flavor changing interactions among fermions of different flavors which
sit at points finite distances apart along the extra dimentions.
These effects come on top of the flavor mixing effects generated
during the electroweak gauge symmetry breaking by the trilinear Yukawa
couplings of fermions to Higgs bosons.  We focus here on a restricted
set of hadronic and leptonic flavor observables believed to be among
the most sensitive ones.  To simplify calculations, we introduce
certain assumptions on the flavor structure of the four fermion
amplitudes which lead to an approximate factorization of the direct
and indirect flavor changing effects.

The outline of the present work is as follows.  Building up on our
previous work~\cite{chem06}, we present in Section~\ref{sect2} the
tree level four fermion string amplitudes for the high energy
processes of fermion-antifermion annihilation into fermion-antifermion
pairs and fermion pair scattering, $f + \bar f \to f ' + \bar f '$ and
$ f + f ' \to f + f' $.  We next consider an approximate construction
of the contact interactions between pairs of quarks and/or leptons
produced by the decoupling of string excitations.  Finally,
specializing to the Standard Model solution of Cremades et
al.,~\cite{CIM,CIM2}, we present numerical results for the chirality
conserving contact interactions as a function of the string and
compactification mass scales and the parameters describing the
separation of intersection points.  The corrections to the Standard
Model contributions are studied over the admissible parameter space
for the string and compactification mass scales.  For
comprehensiveness, we provide in Appendix~\ref{sappen0} a brief review
of the intersecting $D6$-brane models putting a special emphasis on
the topics relating to the parameterization of the branes intersection
points and the Chan-Paton gauge factors which have been lightly
addressed so far.  The discussion of tree level string amplitudes is
complemented in Appendix~\ref{sectappen1} by a review encompassing
both the intersecting $D6$-brane and $ D3/D7$-brane models aimed at
the two-body processes of fermion-antifermion annihilation into pairs
of gauge bosons and of gauge boson scattering, $ f + \bar f \to \g +
\g $ and $ \g + \g \to \g + \g $.

In Section~\ref{sect3}, we discuss the implications from the indirect
high energy collider tests with a special focus on the Bhabha
scattering differential cross section.  In Section~\ref{sect4}, we
examine the contributions from the flavor dependent four fermion
contact interactions to the hadronic and leptonic flavor changing
observables associated to the mass splitting of quark-antiquark
neutral mesons and the lepton number violating three-body leptonic
decays of the charged leptons.  For all the above applications, we
compare our predictions with experimental data in order to infer lower
bounds on the string mass scale at a fixed ratio of the string to
compactification mass scales.

\section{Tree level string amplitudes in intersecting brane models}
\label{sect2}

We calculate the tree level open string amplitudes for four massless
fermion modes localized at the intersection of $D6$-branes in toroidal
orientifold models.  After quoting in Subsec.~\ref{subsect22} the
general formula for the string amplitudes and discussing its
low-energy representation as infinite sums of pole terms and the
subtraction prescription proposed to construct the contact
interactions, we specialize in Subsec.~\ref{subsect23} to the Standard
Model vacuum solution of Cremades et al.,~\cite{CIM,CIM2} and present
in Subsec.~\ref{subsect24} numerical predictions for the contact
interactions. Several notations are specified in
Appendix~\ref{sappen0} which provides a brief review of intersecting
$D6$-branes models.  All calculations are performed with the
space-time metric signature, $ (-+++)$, using units where, $\a ' = {1/
m_s ^2 } = 1 $, except on certain occasions where $ \a ' $ will be
reinstated.

\subsection{Four fermion  string amplitudes and contact interactions}
\label{subsect22}

The open string amplitudes for four fermions localized at points of
the internal manifold can be calculated most conveniently by means of
the superconformal field theory on the world sheet.  The basic tools
were initially developed for the closed string
orbifolds~\cite{bersh87,dixon87} and refined in several subsequent
works (see~\cite{burwick}, for instance).  The application to the
`twisted' or non-diagonal modes of open string sectors was discussed
later for the case of branes within
branes~\cite{hashi96,gava97,fro099,antbengier01} and of intersecting
branes~\cite{Cvetic:2003ch,owen03,abel03,chem06}.

\subsubsection{String amplitudes of localized fermions}

The general configuration of quantum numbers for the four fermion
processes consist of two incoming conjugate fermion pairs, $f_i (k_1)
+ \bar f_j (k_2) + f '_k (k_3) +\bar f' _l (k_4)$, localized at the
four intersection points, $ X _i,\ X_j,\ X_k,\ X_l \in T ^6 $, of the
four $D6$-brane pairs, $ (D,A), \ (A,B) $ and $ (B,C), \ (C,D) $,
intersecting at the angles, $ \mp \t ^I $ and $\mp \t ^{'I}$.  The
tree level open string amplitudes are obtained from the correlators of
vertex operators inserted at points $ x_1,\ x_2,\ x_3,\ x_4 $ on the
disk surface boundary by integrating over the disk moduli space of the
punctured disk.  Using the invariance under the M\"obius group to set,
$ x_1=0,\ x_2=x,\ x_3=1,\ x_4 = X \to \infty $, one can write the
resulting formula as~\cite{chem06} \bea && \cala ' _{f^4} \equiv
{\cala ( V _{-\t , (D,A) , i , k_1 } (x_1) V_{\t , (A,B) ,j , k_2
}(x_2) V_{-\t ' , (B,C) ,k , k_3 }(x_3) V_{\t ', (C,D ) , l , k_4 }
(x_4)) / [i( 2 \pi )^4 \d ^{4} (\sum _i k_i) ] } \cr && = C \bigg (
\cals _{1234} \calT _{1234} \calv _{1234} (s,t) + \cals _{1324}\calT
_{1324} \calv _{1324}(u,t) + \cals _{1243} \calT _{1243} \calv _{1243}
(s,u) \bigg ) , \label{eqstramp1} \eea where we use the notations:
\bea && \calv _{1234} (s,t) = \int _0 ^1 dx x ^{-\a 's -1 } (1- x )
^{-\a 't -1} \prod _I ( {2 \sin \pi \t ^I \over I_I (x) } ) ^{\ud }
\sum _{cl} Z _{cl} ^{I } (x) , \cr && \calT _{1234} = T _{1234} + T
_{4321} ,\ T _{1234} = Tr(\l _1 \l _2 \l _3 \l _4) ,\ \cals _{1234} =
-\cals _{1432}= ( u_1 ^T \g ^0 \g _\mu u_2) (u_3 ^T \g ^0 \g ^\mu
u_4), \cr && C = 2 \pi g_s \a ' = { g_\mu ^2 \vert L_\mu \vert \over 2
m_s ^2 K_\mu },\ \vert L_\mu \vert = \prod _I L_\mu ^I , \ L_\mu ^I =
[(n_\mu ^I r_1^I) ^2 + (\tilde m_\mu ^I r_2^I) ^2 ] ^\ud , \cr && s =
-(k_1 + k_2 )^2 , \ t= -(k_2 + k_3 )^2 , \ u = -(k_1 + k_3)^2
. \label{eqsxx1} \eea The string amplitude $\cala ' _{f^4}$ in
Eq.~(\ref{eqstramp1}) is built from three reduced (partial) amplitudes
associated to the cyclically inequivalent permutations of the
insertion points. The second and third terms are obtained from the
first term associated with the reference configuration $1234$ by
substituting the mode labels $ 2 \leftrightarrow 3 $ and $ 3
\leftrightarrow 4 $, and modifying the interval of the $x$-integral
from $x \in [0,1] $ to $x \in [1, \infty ] $ and $x \in [-\infty , 0]
$.  Each partial amplitude decomposes into a pair of amplitudes
associated to the direct and reverse orientation permutations of the
labels, corresponding to the substitutions, $ 2 \leftrightarrow 4 $
and $ x \leftrightarrow (1-x) $. The requirement that the world sheet
boundary is embedded in the $ T^2_I $ on closed four-polygons with
sides along the branes $ A B C D $, can be satisfied, in general, only
by a single partial amplitude, the conflict with the target space
embedding forcing the other two to vanish. In our present notational
conventions, only the reference term, $ \calT _{1234}\calv _{1234} $,
survives, while the other two terms, $\calT _{1324} \calv _{1324} ,\
\calT _{1243} \calv _{1243} $, cancel out.

The factorization of the residues of the massless pole terms from
exchange of gauge bosons between fermion pairs into products of three
point current vertices determines the normalization factor as, $ C = 2
\pi g_s \a ' $.  Using the familiar results~\cite{polchb} for the
$D$-branes of type $II$ string theories yields the formula in
Eq.~(\ref{eqstramp1}) expressing $C$ in terms of the gauge coupling
constant $ g_\mu $ of the 4-d gauge theory on the $D6_\mu $-brane and
the volume of the three-cycle $\vert L_\mu \vert $ that it wraps. The
same formula applies to each factor of the complete gauge group.  The
orientifold symmetry is taken into account by the factor $ K_\mu $
which is assigned the value $ K_\mu =1 $ or $ K_\mu =2 $ when the
brane $\mu $ is distinct or coincides with its mirror image,
corresponding to the cases with $U(N)$ and extended $ SO(2N) $ or
$USp(2N)$ gauge symmetries, respectively.  It is important to realize
that the relation between the string theory gauge coupling constants,
$ g_\mu $, and their field theory counterparts which we denote
momentarily by $ g ^{ft}_\mu $, also depends on the way in which the
analogous field theory model is constructed.  An illustrative
discussion of the model dependence is presented in~\cite{KW03}.  For
the moment we express this relationship by the proportionality
relation, $ \eta g^{ft}_ \mu = g_\mu $, involving the real parameter $
\eta $.

The factors $ I _I(x)$ and $Z ^I _{cl} = \sum _{cl} e ^{S ^I_{cl} } $
for each $ T_2^I$ in Eq.~(\ref{eqstramp1}) designate the quantum
(oscillator) and zero mode world sheet instanton contributions to the
correlator of coordinate twist fields in the complex plane of $
T^2_I$.  With the choice of independent pair of cycles, $\calc _A =
(x_1, x_2) , \ \calc _B = (x_2, x_3) $, surrounding the insertion
points along the world sheet boundary, $ x_1, \cdots , x_4$, which map
to the intersection points labeled $ i, j, k, l$ in $ T^2_I$, the
summations in the classical partition function $Z ^I _{cl}$ run over
the large lattice generated by the one-cycles $ L_A^I $ and $ L_B^I$
wrapped by the $D6 _A/D6_B$-branes in $ T_2^I$. Going through a full
circle around the cycles $\calc _A ,\ \calc _B,$ induces the
coordinate fields monodromies \bea && \sqrt 2 \D _{\calc _{A} } X =
2\pi (1-e^{2\pi i \t } ) v_{A} ,\ \sqrt 2 \D _{\calc _{B} } X = 2\pi
(1-e^{2\pi i \t } ) v_{B} , \eea where \bea && v_A = p_A L_A + \d
_{ij} ^ A = ( p_A + \e _{ij} ^A) L_A + d _{ij}^A , \ v_B = p_B L_B +
\d _{jk} ^B = ( p_B + \e _{jk}^B) L_B + d _{jk}^B , \eea with $ p_A,\
p_B\in Z $ denoting the winding numbers.  The 2-d large lattices
generated by the brane pairs in the complex planes of $T^2_I$ are
displaced from the origin by the shifts separating the branes
intersection points, $ \d ^A = \e ^A L_A + d^A, \ \d ^B = \e ^B L_B +
d ^B ,\ [ \Re (L _{A,B} ^\star d ^{A,B} ) =0 ] $ where the real
parameters $ \e ^A , \ \e ^B $ and $d ^A , \ d ^B $ stand for the
longitudinal and transverse components of the shift vectors relative
to branes $ A, \ B$.  Detailed formulas for the functions $ I(x)$ and
$Z ^I_{cl} $ can be found in our previous publication~\cite{chem06}.

In the special case involving equal interbrane angles, $\t ^I = \t
^{'I}$, which corresponds to the parallelogram $ DABC$ with $ D=B$ and
$ A=C$, the string amplitude simplifies to \bea && \cala ' _{f^4} = C
\cals _{1234} \bigg [T_{1234} \int _0^1 dx x ^{-s-1} (1-x) ^{-t-1}
\sum _{cl} Z ^{cl I } _{12,34} (x) \cr && + T_{4321} \int _0^1 dx x
^{-t-1} (1-x) ^{-s-1} \sum _{cl} Z ^{cl I } _{43,21} (x) \bigg ] (
{\sin \pi \t ^I \over F(x) F(1-x) } )^{\ud } ,\eea where \bea && Z
^{cl I } _{12,34} (x) = \vt [ { {\e ^ {A } _{12} } \atop 0 } ] (\tau
_A ) \vt [ { {\e ^{B } _{23} } \atop 0 } ] (\tau _B ) = \sum _{p_A,\
p_B \in Z} e ^{ - {\pi } \sin \pi \t _I [ \vert (p_A + \e ^A _{12} )
L_A ^I + d^ A_{12} \vert ^ 2 {F(1-x ) \over F(x) } +\vert ( p_B+ \e
^{B} _{23} ) L_B^I + d^ B_{23} \vert ^ 2 {F(x) \over F(1-x) } ] } ,
\cr && \tau _A (x) = i \sin (\pi \t ^I ) \vert L_A ^I \vert ^2 {F(1-x)
\over F(x)} , \ \tau _B (x) = i \sin (\pi \t ^I) \vert L_B ^I \vert ^2
{F(x) \over F(1-x) } ,\ F(x) = F (\t , 1-\t ; 1; x) , \label{eqamp4}
\eea with $ F(a,b;c;x) $ denoting the Hypergeometric function and $
\vt [ { \t \atop \phi } ] (\tau ) $ the Jacobi Theta function.  We
have separated out the direct and reverse permutation terms in the
quartic order trace factor, $\calT _{1234} $.  Verifying the equality
of the factors multiplying $T _{1234} $ and $T _{4321} $ provides a
useful check on calculations.

The string amplitudes for four fermions localized at the intersections
of $ Dp / D (p+4)$-branes are derived~\cite{antbengier01} by a similar
method to that used for intersecting branes. The resulting formulas
are detailed in Appendix~\ref{sectappen1} along with the similar
results for the two-body processes, $ f + \bar f \to \g + \g $ and $\g
+ \g \to \g + \g $.

\subsubsection{Low-energy representations}

The low-energy limit of compactified string theories is described by
means of series expansions in powers of $\sqrt s / m_s $ and $ (m_s
r)^{-1} = { M_c \over m_s} $, where $\sqrt s $ stands for the energy
variable and $r=1/M_c$ for the characteristic compactification radius
parameter.  In the flat space limit, $ r \to \infty $, both the
partition function and fermion localization factors in the
$x$-integral of Eq.~(\ref{eqstramp1}) can be set to unity, $\sum _{cl}
Z ^I_{cl} = \sum _{cl} e ^{- S^I_{cl} (x) } \to 1,\ I _I (x) \to
1$. Ignoring momentarily the gauge factor, one can express the string
amplitude in this limit by the formula \bea && B(-s, -t) \equiv {u
\over st} \cals (s,t) = \int _0^1 dx x^{-s-1} (1-x)^{-t-1} , \cr &&
[\cals (s,t) = {\G (1-s ) \G (1-t )\over \G (1-s-t )} , \ B(-s, -t)
\equiv {\G (-s ) \G (-t )\over \G (-s-t ) } = \sum _{n=0} ^ \infty
(-1) ^n { (-t-1) \cdots (-t-n) \over n! (-s +n) } ] \eea where we have
exhibited in the second line the representation of Euler Beta function
in terms of an infinite series of poles from $s$-channel exchange of
the string Regge resonance of masses, $ M_n ^2 = n m_s ^2 , \ [ n \in
Z] $.  At finite $r$, applying the familiar method of analytic
continuation past poles to the $x$-integral~\cite{polchb} with the
factors $ Z_{cl} $ and $ I(x)$ included, produces infinite series of
$s$-channel and $t$-channel pole terms at the squared masses of the
open string momentum and winding modes.  Since for the equal angle
case associated with the brane configuration $DABC$ with $C=A, \ B=D$,
the argument variable $\tau _A (x) \to 0 $ at small $x$, one must
carry beforehand the modular transformation on the Theta function with
modular argument $\tau _A(x) $.  At small $1-x$, the same applies to
the Theta function with modular argument $\tau _B(x)$.

selecting the regions near $ x=0$ in the representation of
Eq.~(\ref{eqamp4}) yields the low-energy expansion of the string
amplitude in the open string compactification modes \bea && \cala ' _
{f^4 ,0} \simeq C \cals _{1234} \bigg ( { T _{1234} \over \vert
L_B\vert } \sum _{p_A, p_B} {\prod _I \d _{AB} ^ {-M ^{I2}_{A _{12} ,B
_{23} } } e ^{2i \pi p_B \e ^{BI} _{23}} \over -s +\sum _{I} M
^{I2}_{A _{12},B _{23}} } + { T _{4321} \over \vert L_C\vert } \sum
_{p_C, p_D} { \prod _I \d _{BC} ^ {-M ^{I2}_{B_{32} , C_{43} } } e
^{2i\pi p_D \e ^{BI}_ {32}} \over -t + \sum _{I} M ^{I2} _{ B_{32} , C
_{43} } } \bigg ) , \eea where \bea && M ^{I2} _{A _{ij} ,B _{jk} }=
\sin ^2 (\pi \t _I) (p_A +\e ^{AI} _{ij} )^2 \vert L^I_A \vert ^2 +
{p_B^2 \over \vert L^I_B \vert ^2 } ,\ \ln \d _{AB} = 2 \psi (1) -
\psi (\t ^I) - \psi (1-\t ^I) ,
\label{eqloexp} \eea 
and $\psi (z)$ is the PolyGamma function.  The squared masses, $ M^2
_{A,B} $, consist of string momentum modes of the sector $(B,B)$ along
the compact directions wrapped by the brane $B$, and string winding
modes of the sector $(A,A)$ for strings stretched transversally to the
cycle $ L_B$.  Expressing the normalization factor $C = 2 \pi g_s$ in
the $s$- and $t$-channel pole terms in terms of the corresponding
parameters of branes $ B$ and $C$, respectively, then comparison with
the amplitude of the analogous field theory leads to the
identifications, $ {g_B^2 \over 2 K_B} T_{1234} = (g_B^{ft })^2 \sum
_a (T^a)_{12} (T^a ) _{34} $ and ${g_C^2 \over 2 K_C} = (g_C^{ft } )^2
\sum _a (T^a)_{23} (T^a) _{14} $, where $a$ labels the Lie algebra
generators of the gauge groups on branes $B,\ C$.  The extra factor
$2$ in the denominator accounts for the field theory normalization
convention used for the field theory gauge coupling constant,
corresponding to the gauge current vertex, $ < J ^a _\l (0) > = g_A
T^a \g _\l , \ [Trace(T_a T_b) =\ud \d _{ab}]$.

The residues $ \d _{AB} ^{-M _{A,B}^2}$ of the poles at $s = M^2
_{A,B} $ (and the similar residues of the poles at $ t = M^2 _{B,C} $)
represent the squares of the form factors for the three point
couplings of fermion pairs localized at the $D6_A/ D6_B$ brane
intersections with the open string states of mass $M _{A,B} $ from the
sectors $(A,A)$ and $(B,B)$.  As the interbrane angle runs over the
defining interval, $\t ^I\in [0, 1]$ in the $ T^2_I$ tori, $\ln \d ^I
_{AB}$ asymptotes to $ \cot (\pi \t ^I) \to +\infty $ at the interval
end points while reaching the minimum value, $\ln \d ^I _{AB} \simeq
2.77 $, at the mid-point $ \t ^I = \ud $.  The divergence of $\ln \d
^I _{AB}$ at $\t ^I=0 $ and $ 1 $ implies then the absence of
contributions from the exchange of string compactification modes for
parallel branes, as expected by virtue of the momentum conservation.
The form factor, $ \calf _{AB} (p_B) = \d _{AB} ^{-M _{A,B}^2 /2 }$,
representing the cost for a fermion particle to absorb the gauge boson
momentum $ p_B$, arises as a consequence of the $D$-brane fuzziness
caused by the finite spatial extension of string modes.  Its
configuration space representation, $\calf_{AB} (y) $, can be
evaluated by writing the three point coupling of fermion pairs to the
momentum modes of the $ D6_B$-brane gauge connection field, $A ^B_{\mu
} (x,y)$, as a Fourier integral over the cycle of radius $ L_B$ \bea
&& \int _0 ^ {\pi \vert L_B\vert } dy A ^B_{\mu } (x,y) \calf_{AB} (y
-y_i^B) \equiv \sqrt {2\over \pi \vert L_B\vert } \sum _{p_B} N_{p_B}
[ A ^{B(p_B)} _{\mu } (x) \int _0 ^ {\pi \vert L_B\vert } dy \cos ({
yp_B\over \vert L_B\vert }) \calf_{AB} (y -y_i^B) +\cdots ] \cr && =
\sqrt {2\over \pi \vert L_B \vert } \sum _{p_B} N_{p_B} [ A ^{B(p_B)}
_{\mu } (x) \d _{AB} ^{- { p_B ^2 / (2 \vert L_B \vert ^ 2 ) } } e ^{i
p_B y_i^B \over \vert L_B\vert } +\cdots ] , \eea with the resulting
formula \bea && \calf _{AB} (y) \simeq \sqrt{ 2 \over \pi \ln \d _{AB}
} e ^{- {y ^2 \over 2 \vert L_B \vert ^ 2 \ln \d _{AB} } } ,\ [y_i^B =
2 \pi \e _i ^B \vert L_B\vert ]
\label{eqformf} \eea where $ y $ parameterizes the points on the $
L_B$ cycle which has been described here by the orbifold $ S^1/Z_2$ of
length $ \pi \vert L_B\vert $; $y_i ^B $ denotes the position of the
fermion mode; $ N_{p_B}= 1 $ for $ p_B\ne 0 $ and $ N_{p_B}= 1/\sqrt 2
$ for $ p_B =0 $; and the central dots stand for the sine stationary
modes.  The above approximate formula for $\calf _{AB} (y) $ becomes
exact in the large radius limit, $\vert L_B \vert \to \infty $.  The
exchange of massive KK gauge bosons contributes to the four fermion
amplitude a sum of pole terms with residue factors, $ e ^{2\pi i p _B
\e ^B _{jk} } \d _{AB} ^{- p_B ^2 /\vert L_B \vert ^2 } $, where $\e
^B _{jk} $ denotes the relative distance along the brane $B$, as
expected by comparison with Eq.~(\ref{eqloexp}). A similar analysis
holds for the momentum modes associated with the brane $C$.  By
contrast, there is no field theory interpretation for the form factors
accompanying the coupling of localized fermions to the open string
winding modes.

To illustrate further how the form factor originates within the field
theory framework, we consider the toy-like 5-d $U(1)$ gauge theory
with the fifth dimension compactified along the orbifold segment, $ y
\in [0, \pi \vert L _B \vert ]$, assuming that the chiral fermions are
trapped near the boundaries by some soliton kink solution involving a
scalar field coupled to the fermions.  Making use of the Gaussian
ansatz for the normalizable zero mode wave function of a fermion
localized at $ y_i^B$, \bea && \psi ^{(0)}_i (y) = N e ^{ - (y-y_i^B)
^2 / (2 \s ^2 \pi ^2) } ,\ [\int _0 ^{\pi \vert L _B\vert } dy \psi
^{(0) \star }_i (y) \psi ^{(0)}_i (y) =1 , \ N \simeq ( {2 \over \pi
^{3\over 2} \s } ) ^\ud ] \eea where the normalization integral
determining $ N$ has been evaluated in the limit $ L_B \to \infty $,
we infer the gauge vertex coupling \bea && \sum _{p_B} N_{p_B} A_\mu
^{B (p_B)} (x) \int _0^ {\pi \vert L _B\vert } dy \psi ^{(0)\star }_i
(y) \psi ^{(0)}_i (y) \cos ({y p_B \over\vert L _B\vert }) \simeq
\sum _{p_B} N_{p_B} e ^{- { \pi ^2 \s ^2 p_B^2 \over 4 \vert L_B\vert
^2 } } \cos ({ y_i^B p_B \over \vert L_B \vert } ) . \eea Comparison
with the form factor $\calf _{AB} (p_B)$ in Eq.~(\ref{eqformf}) allows
us to identify the half-width parameter as, $ \s \simeq \sqrt { 2 \ln
\d _{AB} } / \pi $.  The above result for the form factor of fermion
modes localized at intersecting branes agrees with that
derived~\cite{antbengier01} for the fermion modes of the non-diagonal
sectors of $D3/D7$-branes, where $\t ^I =\ud $.  (The half-width
parameter in~\cite{antbengier01}, which we distinguish here by the
suffix label $ {ABL} $, is related to ours as, $ \s _{ABL} = \s \pi
/\sqrt 2 $. For later reference, we note that our half-width parameter
$ \s $ identifies with the parameter denoted $ \s $ in~\cite{lill03}.)
It is of interest to note that a similar form factor also arises in
the string amplitude for emission of a massive graviton mode $ G
(M_{cl})$ in the two-body reaction, $ f + \bar f \to \g + G (M_{cl})$,
with the characteristic dependence on the mass $M_{cl}$ of the closed
string mode~\cite{cpp00}, $ \calf _{G} (M_{cl}) = e ^{-\ln (2 )
M_{cl}^2 /m_s^2 } $.  Both the present string form factor and the one
derived above, $ \calf _{op} (M_{op}) = e ^{- M^2 _{op} \ln (\d ) /
(2m_s^2 )} $, should be distinguished from the form factor arising
from the quantum fluctuations of
branes~\cite{bando99,hisano99,murawells01}.  Using a quantum field
theory treatment of the Nambu-Goto action for domain walls with the
transverse coordinate $y$ promoted to the would-be light
Nambu-Goldstone scalar field associated to the spontaneous breaking of
translational invariance, Bando et al.,~\cite{bando99} found that the
coupling to momentum modes in the effective 4-d action is accompanied
by the recoil form factor \bea && \calf _{rec} (M) \simeq e ^{ - \D ^2
M ^2 } , \ [\D ^2 = {M_F ^2 \over (4 \pi ) ^2 \tau _{p} } = {M_F ^2 g
^2 _{Dp} \over 4 m_s ^4} ,\ M ^2 = {p^2 \over \vert L_B\vert ^2 }]
\eea where $ M_F $ denotes the ultraviolet mass cut-off for the field
theory on the domain wall whose tension parameter has been identified
above to that of the $ Dp$-brane, $ \tau _{Dp} = [m_s ^2 / (2 \pi
g_{Dp}) ]^2 $, with gauge coupling constant $g _{Dp}$.  The specific
case of a soliton domain wall is discussed in ~\cite{hisano99}.  From
the comparison of the ratio of half-width parameters for the form
factors associated to the brane fuzziness and recoil, $\D ^2:\s ^2 =
{M_F^2 g^2 _{Dp} \over 4m_s^4 } : {\ln \d \over 2m_s^2} \simeq {g^2
_{Dp} \over 2\ln \d } : 1 $, where we have assumed in the second
stage, $ M_F \simeq m_s $, we conclude that the suppression effect
from brane recoil is parametrically weaker than that from the string
finite size. This property was previously discussed in~\cite{cpp00}.

\subsubsection{Contact interactions and flavor structure}

The four fermion string amplitude receives infrared divergent
contributions from the $x$-integral boundaries which correspond to the
massless $s , \ t $ channel pole terms from gauge bosons exchange.
One way to regularize these harmless poles is to subtract out the
small regions in the $x$-integral near the end points $ x=0 ,\ x=1 $,
while adding up the corresponding pole terms by hand, as described by
the subtraction, $ \cala \to \cala - (- \cala _s^{pole} + \cala \vert
_{x=0} ) - ( -\cala _t^{pole} + \cala \vert _{x=1} ) $.  To account
for the electroweak symmetry breaking, one can use the same
prescription where the added pole terms correspond to the
contributions from exchange of the physical gauge bosons with the
observed finite values of the masses. For the four fermion coupling
$(\bar f_H \g _\mu f_H ) (\bar f'_{H'} \g ^\mu f'_{H'} ) $ in the
electrically charge neutral channel for the $ \g , \ Z$ gauge bosons,
this is illustrated by the substitution, $ {1 \over s} \to {1 \over s}
+ { a_H (f) a_{H'} (f') \over s -m_Z ^2 } $, where $ a_H(f)$ denote
the $Z$ boson vertex couplings.  A similar replacement holds for the
$t-$channel poles.

The contact interactions, subsuming the contributions from the massive
string Regge and winding excitations, can be constructed in a similar
way by subtracting also the pole terms associated to the string
momentum excitations, as illustrated by the schematic formula \bea &&
\cala '_{contact} = \cala '- \cala '_{s, KK } - \cala' _{t, KK}
. \label{eqsubtr} \eea

The same subtraction procedure was previously used to
define~\cite{antbengier01} the contact interactions in models with $
Dp/D(p+4)$-branes.  For consistency, we remove the $s-$channel terms
only for the configurations of intersection points with $ i = j $ or
$\e ^A _{ij} = 0,$ and the $t-$channel terms only for those with $k =
l $ or $ \e ^B _{jk} = 0$.  No subtraction of momentum modes is needed
in the cases, $ i \ne j \ne k $ or $\e ^A _{ij} \ne 0 , \ \e ^B _{jk}
\ne 0$, where the thresholds for the momentum modes are separated by a
finite gap corresponding to the mass terms, $ \e _A ^2 \vert L_A\vert
^2$.  One motivation for excluding the KK towers from the contact
interactions is simply that it is always possible to include
separately their contributions through the familiar field theory
treatment, suitably generalized by the inclusion of form factors.  In
the flat space limit, $ L _B\to \infty $, with the brane form factors
set to unity, $ \d ^{- M_{A,B}^2} \to 1 $, the contact interactions
from KK modes are described by the approximate formula \bea && L_{EFF}
= { g_B^2 \over 2 m_s ^2 } \cals _{1234} \calT _{1234} \sum _{p \in Z}
{\d ^{- p^2 / \vert L_B \vert ^2} \over ( p^2 / \vert L_B \vert ^2 -s
) } \simeq { g_B^2 \over 2 m_s ^2 } { S_n (m_s \vert L_B \vert ) ^n
\over ( n-2) } \cals _{1234} \calT _{1234} , \label{eqcontkk} \eea
where $ S_n = {2 \pi ^{n/2} \over \G ( n/2) } $ with $ n = D-4$
denoting the number of real dimensions of the wrapped cycle.

In the kinematic regime where mass parameters and energies are
negligible in comparison to $ m_s $, the four fermion local couplings
are dominated by the chirality conserving couplings of dimension
$\cddd =6$. The corresponding terms in the Standard Model effective
Lagrangian are represented in the left and right chirality basis of
fermion field operators by the general structure \bea && L _{EFF} =
\sum _{f , f'} [G _{ff'}^{LL} (\bar f _L \g _\mu f _L ) (\bar f' _L \g
^\mu f' _L ) + G _{ff'}^{LR} (\bar f _L \g _\mu f_L ) (\bar f '_R \g
^\mu f '_R ) \cr && + G _{ff'}^{RL} (\bar f _R \g _\mu f _R ) (\bar f'
_L \g ^\mu f ' _L ) + G _{ff'}^{RR} (\bar f _R \g _\mu f _R) (\bar f'
_R \g ^\mu f' _R ) ] , \label{eq4fernions}\eea where $ G _{ff'}^{ HH'}
= \eta _{ HH' } { 4 \pi \over 2 \L ^{ ff' 2 }_{ HH' } } $, with $ H, \
H' = (L, R) $.  We follow the notational conventions for the
coefficients $G _{ff'}^{HH'} $ commonly adopted in the context of
composite models~\cite{peskin83}, with $\eta _{ HH'} $ denoting
relative $\pm $ signs and $ \L ^{ ff' 2 }_{ HH' } $ the characteristic
squared mass scale parameters.  For later reference, we also quote the
effective Lagrangian of use in leptonic collider applications \bea &&
L _{EFF} = \sum _{f, H, H'} { 4 \pi \eta _{ HH' } \over \L ^{ ef 2 }_{
HH' } (1 +\d _{e f} ) } G _{ef}^{HH'} (\bar e _H \g _\mu e _H ) (\bar
f _{H'} \g ^\mu f _ {H'} ),\
\label{eq4electrons}\eea where $ \d _{e f}= 1 $ if $e=f$ and $ \d = 0
$ if $e \ne f$.
The left chirality basis for the fermion fields is related to the
mixed left-right chirality basis by, $ L \sim R^ {c \dagger } $, thus
implying the identity, $(\bar L_1\g ^\mu L_2) = (\bar R_2 ^c \g ^\mu
R_1^c) $, and the Fierz-Michel identities for the quartic order matrix
elements, \bea && (\bar L_1\g ^\mu L_2)(\bar R_3\g _\mu R_4) = 2 (\bar
L_1R_4)(\bar R_3L_2), \ (\bar L_1\g ^\mu L_2)(\bar L_3\g _\mu L_4) = -
(\bar L_1\g ^\mu L_4)(\bar L_3\g _\mu L_2) ,\eea where $ L,\ R $
denote c-number (commuting) Dirac spinors.  The comparison with
Eq.~(\ref{eqstramp1}) yields the explicit formula for the
dimensionless coefficient functions $ g _{ff'} ^{HH'} (L_\mu )$
defined by \bea && g _{ff'} ^{HH'} \equiv { m_s ^2 K_\mu \over 2 \pi
g_\mu ^2 } G _{ff'}^{HH'} = \eta _{ HH'} { m_s ^2 K_\mu \over \L ^{ff'
2} _{HH'} g_\mu ^2 } = {\vert L_\mu \vert \over 4 \pi } \int _0 ^1 dx
x^{-s -1} (1-x)^{-t -1} \prod _I ( {2\sin \pi \t ^I \over I_I(x) }
)^{\ud } \sum _{cl} Z ^I _{cl } ,
\label{eqcoefg} \eea such that the string  four fermion amplitude
reads as, $ \cala _{f^4} = {2 \pi g^2 _\mu \over m_s ^2 K_\mu }
g_{ff'} ^{HH'} <\cals _{1234} > _{HH'} <\calT _{1234} > _{ff'}.$ For
convenience, we shall employ in the sequel a similar notational
convention for the string amplitudes for fermions of fixed
chiralities.

To make contact with the amplitudes of physical processes involving
the mass eigenstate fields, we need to perform the familiar bi-unitary
linear transformations linking the above gauge basis to the mass
eigenvector basis, $ f \to V _L ^{f \dagger } f, \ f ^c \to V_R ^{f T
} f ^c ,\ [f=q, l] $ which read in the left-right chirality basis as,
$ f _H \to V _ H^{f \dagger } f_H ,\ [H=L, R]$. The flavor mixing
matrices, $ V ^f_L,\ V ^f _R $, are determined through the
diagonalization of the fermion mass matrices in generation space, $
V_R ^f M_f V_L ^{f\dagger } = (M_f)_{diag}$, but in a partial way
since the Standard Model contributions from the quarks and leptons
only depend on the products, $ V_{CKM} = V_L ^u V_L ^{d \dagger } $
and $ V' = V_L ^\nu V_L ^{e \dagger } $, where the suffix label CKM
refers to the quarks Cabibbo-Kobayashi-Maskawa matrix. The effective
Lagrangian of dimension $\cddd =6$ in the vector spaces of the
fermions generation and mass basis fields can now be expressed as \bea
&& L_{EFF} =\sum_{f,f'} \sum_{H,H'} G^{HH'} _{ij,kl} [(\bar f_{iH} \g
^\mu f_{jH}) (\bar f'_{kH'} \g _\mu f'_{lH'})]_{flav} +H.c.  \cr &&
=\sum_{f,f'} \sum_{H,H'} \tilde G^{HH'} _{ij,kl} [(\bar f_{iH} \g ^\mu
f_{jH}) (\bar f '_{kH' } \g _\mu f'_{lH'})]_{mass} +H.c., \eea where
\bea && \tilde G^{HH'} _{ij,kl} \equiv { 2 \pi g_A ^2 \over m_s ^2
K_A} \tilde g _{ij, kl} ^{HH'} = G^{HH'} _{i'j',k'l'} (V^{f} _{H ii'}
V^{f\star } _{Hjj'} ) (V^{ f'} _{H 'kk'} V^{f' \star } _{H 'll'} ) ,\
G^{HH'} _{ij,kl} \equiv { 2 \pi g_A ^2 \over m_s ^2 K_A} g _{ij,kl}
^{HH'} \eea by using the familiar tensorial notation for the flavor
and mass bases coefficients in Eqs.~(\ref{eqcoefg}) which are labeled
by the same indices $ i, j, k, l \in [1,2,3]$.  The $4$-point
couplings $ G^{HH'} _{ij, kl} $ of localized modes are subject to
geometrical selection rules which are expressed in terms of the shift
vectors, $ w ^I_{BA} $, associated to the embedding polygon with sides
$ D, A, B, C $ in each $T^2_I $, by the conditions~\cite{higaki05}
\bea && w ^I_{DA} + w ^I_{AB} + w ^I_{BC} + w ^I_{CA} = 0 \ \text{mod
}\ \L ^I,\ [w ^I_{BA} \in \L ^I _{BA} / \L ^I ],
\label{eqgeomsel} \eea  where   we use notations  defined just below Eq.~(\ref{eqshifts}).


\subsection{ Standard Model realization with four branes}
\label{subsect23}

We here specialize to the solution of Cremades et al.,~\cite{CIM,CIM2}
with the brane setup consisting of four $D6$-brane (baryon, left,
right and lepton) stacks of size $ N_a=3,\ N_b=1,\ N_c =1,\ N_d=1$,
supporting the extended Standard Model gauge symmetry group, $ U(3)_a
\times SU(2)_b \times U(1) _c \times U(1) _d$.  The weak gauge group,
$ USp(2)_b\sim SU(2)_b$, identifies with the enhanced gauge symmetry
associated with the overlapping pair of mirror $ D6_b/D6_ {b'}
$-branes, and the hypercharge with the linear combination of Abelian
charges, $Y = {Q_a\over 6} - {Q_c\over 2} - {Q_d\over 2} $.  As noted
by Kokorelis~\cite{kokosusy02}, the present brane setup belongs to the
family of solutions described by the winding numbers and intersection
angles listed in the following table whose members are labeled by the
discrete parameters, $\rho = (1,\ 1/3 ), \ \e =\pm 1 , \ \tilde \e
=\pm 1 , \ [\e \tilde \e =1] $ with $ \b _i =1- \hat b_i =[1, \ud ] ,
\ [i=1,2] $ in correspondence with the cases of orthogonal and flipped
tori $ T^2_{2,3}$.
\vskip 0.5 cm
\begin{center}
\begin{tabular}{c|ccccc} 
Brane $ (N_\mu )$ & $ (n ^1_\mu , m ^1_\mu ) $ & $ (n ^2_\mu , m^2_\mu
) $ & $ (n ^3_\mu , m^3_\mu ) $ & $ \t _\mu ^{I=1,2,3} $ & Susy
Charges \\ \hline &&&&& \\ Baryon $ (N_a=3)$ & $ (1,0) $ & $ ({1\over
\rho } ,3\rho \e \b _1 ) $ & $ ({1\over \rho } ,-3\rho \tilde \e \b _2
) $ & $ (0, \pm \t _a, \mp \t _a) $&$ {r_{(1)} , r_{(4)}} $\\ &&&&& \\
Left $(N_b=1)$ & $ (0,\e \tilde \e ) $ & $ ({1\over \b _1} ,0) $ & $
(0,-\tilde \e ) $ & $ (\t _b ,0, \mp \t _b ) $&$ {r_{(1)} , r_{(3)}
\choose r_ {(2)}, r_ {(4)}} $ \\ &&&&& \\ Right $(N_c=1)$ & $ (0,\e )
$ & $ (0,-\e ) $ & $ ( {\tilde \e \over \b _2} ,0) $ & $ (\pm \t _c,
\mp \t _c ,0) $&$ {r_{(3)}, r_{(4)}} $ \\ &&&&& \\ Lepton $ (N_d=1)$ &
$ (1,0) $ & $ ({1\over \rho } ,3\rho \e \b _1) $ & $ ({1\over \rho }
,-3\rho \tilde \e \b _2 )$ & $ (0,\pm \t _d, \mp \t _d ) $ & $
{r_{(1)}, r_{(4)}} $\\
\end{tabular}
\end{center} \vskip 0.5 cm  
  The massless spectrum of left chirality multiplets localized at the
intersection points includes three generations of quarks and
leptons. For the specific choice of brane angles characterized by a
single vanishing angle and a pair of angles equal up to a sign, the
various brane pairs preserve $ \caln = 2$ supersymmetry each, provided
the complex structure parameters of $ T^2 _{2, 3}$ satisfy the
relation, $ \b _1 \chi _2 = \b _2 \chi _3$.  We have indicated in the
last two columns of the above table the brane-orientifold angles and
the spinor weights of the conserved supercharges, with the upper and
lower signs corresponding to $\e =\tilde \e = \pm 1$.  The finite
intersection angles can then be expressed as, $\t _b = \t _c = \ud , \
\t _a = \t _d = {1\over \pi } \tan ^{-1} (3 \rho ^2 \e \b _1 \chi _2)
$.  Only when $ \e = \tilde \e = -1$ do all the four branes share the
common spinor supercharge, $ r_{(4)} = (----)$, implying the existence
of an unbroken $\caln =1$ supersymmetry in this case.  The assignment
of open string sectors and gauge group representations for the quarks
and leptons and for the two Higgs bosons is displayed in the following
table in correspondence with the gauge group $ SU(3)_a\times
SU(2)_b\times U(1) _a \times U(1) _c \times U(1) _d $.
\vskip 0.5 cm
\begin{center}
\begin{tabular}{c | c c c c c c | c c} 
Mode & $ q $ & $ u^c $ & $ d^c $ & $ l $ & $ e^c $& $ \nu ^c $ & $ H_d
$ & $ H_u $ \\ &&&&&&&& \\ \hline &&&&&&&& \\ Brane & {\small $ (a,b)
+ (a,b') $ } & $ (c,a) $ & $ (c',a ) $ & {\small $ (d,b)+(d,b') $} & $
(c',d) $ &$(d,c ) $ & {\small $ (c,b)+ (c,b ')$ } & {\small $(c,b)
^\dagger + (c,b') ^\dagger $ } \\ &&&&&&&& \\ Irrep & $3(3,2)_{1,0,0}
$ & $ 3(\bar 3,1)_{-1,1,0} $ & $ 3(\bar 3,1)_{-1,-1,0}$ & $
3(1,2)_{0,0,1}$ & $ 3(1,1) _{0,-1,-1}$& $ 3(1,1) _{0,-1,1}$& $ (1,2)
_{0,1,0}$ & $ (1,2) _{0,-1,0}$ \\ &&&&&&&& \\
\end{tabular}
\end{center} \vskip 0.5 cm 
  The massless scalar modes for the pair of up and down Higg bosons
arise as the hypermultiplet of the sector $(c,b) = (c,b')$ with $
\caln =2$ supersymmetry, due to the coincidence of the branes $c$ and
$ b$ along the first complex plane $ T^2_1$.  The effective gauge
field theory for this family of models is free from anomalies, despite
the fact that the brane setup fails to satisfy the RR tadpole
cancellation conditions. These can be satisfied, however, by including
a hidden sector of distant brane stacks which do not intersect with
the observable brane stacks.  The quark generations~\cite{CIM,CIM2}
are located in the three complex planes of $ T^2 _I$ at the
intersection points \bea &&
P_{ab} ^{(k)} (q) = \pmatrix{\e _q^{(1)} & -{k\over 3} & \e_q ^{(3)}
\cr \tilde \e _q^{(1)} & 0 & 1 - 3 \e_q ^{(3)} },\ {P_{ca} ^{(k)}
(u^c)\choose P_{c'a} ^{(k)} (d^c)} = \pmatrix{ \e_q ^{(1)} +r_{bc} &
\e_q ^{(2)} & {k\over 3} \mp {\tilde \e_q ^{(3)} \over 3 } \cr \tilde
\e_q ^{(1)} & 3 \e_q ^{(2)} & \pm \tilde \e_q ^{(3)} },
\eea where the three entries in the upper and lower arrays stand for
the coordinates in the orthogonal reference frames, $\Re (X^I) /( 2\pi
r_1^I ) $ and $\Im (X^I) /( 2\pi r_2^I ) ,\ [I=1,2,3 ] $.  The
intersection points for the lepton generations, $P_{db}^{(k)} (l) ,\
P_{c'd} ^{(k)} (e^c),\ P_{dc} ^{(k)} (\nu ^c) $, are described by
similar formulas to those for the quark generations with $ \e ^{(I)}
_{q} \to \e ^{(I)} _{l},\ \tilde \e ^{(I)} _{q} \to \tilde \e ^{(I)}
_{l}.$ The intersection points along the branes are labeled by the
integer index $ k \in [0, 1, 2] \simeq [0, 1, -1] $, with the branes
transverse distance described by the real parameters, $ \e ^{(1)} _{q,
l} , \ \tilde \e ^{(1)} _{q, l} , \ \e ^{(2)} _{q, l}, \ \e ^{(3)}
_{q, l} , \ \tilde \e ^{(3)} _{q, l} , \ r_{bc}$.  (We deviate only in
the definition of the parameters in the third complex plane with the
notations of~\cite{CIM,CIM2} which use the choice of unit of length,
$\Im (X^{I=3} ) /( 6\pi r_2^{I=3}) $. The relationship between us and
them reads, $ (\tilde \e_q ^{(3)} )_{us} = 3 (\tilde \e_q ^{(3)}
)_{them} $.)  Numerical studies for subsets of these parameters have
been reported~\cite{abel03,abeleb03} in attemps to fit the quarks and
leptons Yukawa coupling constant matrices.  Directions to improve
certain shortcomings of the predictions for the fermions mass matrices
are reviewed in~\cite{chamoun03}.  Two important features of the
present family of models are that the intersection points are
separated along the branes by distances of order, $ \e \sim 1/3 $, and
that all three intersection points for the electroweak doublet and
singlet fermion modes are placed at points which lie at finite
distances apart only in the single complex planes, $ T^2_3$ and $
T^2_2$, respectively.
 
The gauge matrices are constructed along the lines traced out in
Appendix~\ref{sappen0}.  To the electroweak singlet leptons, $ e ^c =
(c',d) ,\ \nu ^c = (d,c)$, with multiplicities, $ I _{c'd} = -3,\ I
_{cd} = 3 $ and $ U(1)_{c} \times U(1)_{d} $ charges, $ (Q_c,\ Q_d ) =
(-1,-1)$ and $ (+1, -1) $, we ascribe in the subspace of gauge quantum
numbers $ (c,d) $ the matrices $\l ^{(cd)} _{ e ^c } ,\ \l ^{(cd)}
_{\nu ^c } $ with non-vanishing entries $\g $ and $\a $, respectively,
using the notations of Eq.~(\ref{eqcp}).  Explicitly, $ \g =1 $ for $
e^c \sim (c',d ) _{-1,-1} $ and $ \a =1 $ for $\nu ^c \sim (d,c )
_{-1,1} $, with other entries set to zero.  For the modes charged
under the color group $ SU(3)_a $, choosing the subspace $ (a,c) $, we
ascribe to the electroweak singlet quarks, $ u^{c\dagger } \sim (a,c),
\ d^{c\dagger } \sim (a,c') $, of charges $ (Q_a,\ Q_c)= (+1, \mp 1)
$, the bifundamental representation matrices $ \l ^{(ac)}
_{u^{c\dagger } } ,\ \l ^{(ac)} _{d^{c\dagger } } $ with the
non-vanishing entries for the column and array vectors, $\a _\a ,\ \g
_\a $, in the notations of Eq.~(\ref{eqcp}), transforming under the $
SU(3)$ group fundamental representations.  For the electroweak gauge
group, $ USp(2) \sim SU(2) $, supported by the $D6_b$-brane with the
$USp$ projection, the two components of the electroweak doublet lepton
mode, $ l = (\nu \ e ) ^T \sim (d,b) $, with charges, $ (Q_d, T_3^b) =
(1, \pm 1)$, are ascribed the matrices, $\l ^{(bd)} _{l} $, with the
non-vanishing entries, $ \g $ and $ \a $, using the notations of
Eq.~(\ref{eqcp}).  A similar construction holds for the two components
of the electroweak doublet quark mode, $ q=(u, \ d)$, which are
ascribed the matrices $\l ^{(ab)} _q $, with the non-vanishing
entries, $ \b _\a $ and $ \a _\a $.  The CP matrices of the quarks and
leptons modes are then given by the explicit formulas \bea && \l
^{(cd)} _{e^c} = { 1 \over \sqrt 2 } \pmatrix{0 & 0 & 0 & 0 \cr 0 & 0
& 1 & 0 \cr 0 & 0 & 0 & 0 \cr 1 & 0 & 0 & 0 } , \ \l ^{(cd)} _{\nu ^c}
= { 1 \over \sqrt 2 } \pmatrix{0 & 0 & 1 & 0 \cr 0 & 0 & 0 & 0 \cr 0 &
0 & 0 & 0 \cr 0 & -1 & 0 & 0 } , \cr && \l ^{(db)} _{e } = { 1 \over
\sqrt 2 } \pmatrix{0 & 0 & 1 & 0 \cr 0 & 0 & 0 & 0 \cr 0 & 0 & 0 & 0
\cr 0 & -1 & 0 & 0 } , \ \l ^{(db)} _{\nu } = { 1 \over \sqrt 2 }
\pmatrix{0 & 0 &0 & 1 \cr 0 & 0 & 0 & 0 \cr 0 & 1 & 0 & 0 \cr 0 & 0 &
0 & 0 } , \cr && \l ^{(ac)} _{u^c_\a } = { 1 \over \sqrt 2 }
\pmatrix{0 & 0 & 0 & 0 \cr 0 & 0 & 0 & - \d _\a ^\star \cr \d _\a
^\dagger & 0 & 0 & 0 \cr 0 & 0 & 0 & 0 } , \ \l ^{(ac)} _{d^c _\a } =
{ 1 \over \sqrt 2 } \pmatrix{0 & 0 & 0 & 0 \cr 0 & 0 & \g _\a ^\star &
0 \cr 0 & 0 & 0 & 0 \cr - \g _\a ^\dagger & 0 & 0 & 0 } , \cr && \l
^{(ab)}_{u_\a } = { 1 \over \sqrt 2 } \pmatrix{0 & 0 & 0 & \b _\a \cr
0 & 0 & 0 & 0 \cr 0 & \b _\a ^T & 0 & 0 \cr 0 & 0 & 0 & 0 } , \ \l
^{(ab)}_{d_\a } = { 1 \over \sqrt 2 } \pmatrix{0 & 0 & \a _\a & 0 \cr
0 & 0 & 0 & 0 \cr 0 & 0 & 0 & 0 \cr 0 & -\a _\a & 0 & 0 } , \eea where
the normalization condition is of form, $ Tr(\l _{f _\a } \l _{f _\b }
^ \dagger ) = \d _{ \a \b } $, and, for convenience, we have kept a
record of the pair of branes vector subspaces associated to each mode.
The quartic traces of CP matrices are easily calculated from the above
explicit representations.  The sums of traces for the direct and
reverse orientation terms, indicated below by the suffix label $ (d+r)
$, are given by the formulas \bea && Tr _{(d+r)} (\l ^\dagger _{(e)}
\l _{(e)} \l ^\dagger _{(e)} \l _{(e)} ) = Tr _{(d+r)} (\l ^\dagger
_{(e^c)} \l _{(e^c)} \l ^\dagger _{(e^c)} \l _{(e^c)} ) = 1 , \ Tr
_{(d+r)} (\l ^\dagger _{(e^c)} \l _{(e^c)} \l _{(e)} \l ^\dagger
_{(e)} ) = \ud , \cr && Tr_{(d+r)} (\l ^\dagger _{(q _\a )} \l _{(q
_\b )} \l ^\dagger _{(e )} \l _{( e )} ) =Tr_{(d+r)} (\l _{(d ^c _\a
)} \l ^\dagger _{(d^c _\b )} \l _{( e^c )} \l ^\dagger _{(e^c )} )
=Tr_{(d+r)} (\l _{(u ^c _\a )} \l ^\dagger _{(u^c _\b )} \l ^\dagger
_{(e^c )} \l _{( e^c )} ) = {1 \over 2} \d _{ \a \b } , \cr &&
Tr_{(d+r)} (\l _{(q^c _\a )} \l ^\dagger _{(q^c _\b )} \l _{(q^c _\g
)} \l ^\dagger _{(q^c _\d )} ) = Tr_{(d+r)} (\l _{(q _\a )} \l
^\dagger _{(q _\b )} \l _{(q _\g )} \l ^\dagger _{(q _\d )} ) = T_{ \a
\b \g \d } , \cr && \ Tr_{(d+r)} (\l _{(q _\a )} \l ^\dagger _{(q _\b
)} \l ^\dagger _{(q ^c _\g )} \l _{(q ^c_\d )} ) = {1\over 2 } T_{ \a
\b \g \d } ,\ [T_{ \a \b \g \d } = \ud (\d _{ \a \b } \d _{ \g \d } +
\d _{ \a \d } \d _{ \g \b } ) , \ q=u,d] .  \label{eqtracecp} \eea
Making use of the $U(N_c)$ group identity, $ 2 \sum _{a=1}^{N_c^2-1}
(T^a) _{\a \b } (T^a) _{ \g \d } = (\d _{\a \d } \d _{ \b \g }
-{1\over N_c} \d _{\a \b } \d _{ \g \d } ) $, with the first and
second terms inside parentheses being associated to the non-Abelian
and Abelian group factors $SU(N_c)$ and $U(1)$ of $U(N_c)$, one can
cast the tensorial structure of the trace factor for quarks in the
fundamental representation of $ SU(N_c)$ into the operator form \bea
&&\calT _{1234} = T_{\a \b \g \d } \equiv \ud (\d _{\a \b } \d _{ \g
\d } + \d _{\a \d } \d _{ \b \g } ) = \sum _{a=1}^{N^2 _c-1} (T^a)
_{\a \b } (T^a) _{ \g \d } + {N _c+1\over 2 N_c} \d _{\a \b }\d _{ \g
\d } , \eea where the traceless Lie algebra generators of $SU(N_c)$
are normalized as, $ Trace(T_a T_b) = \ud \d _{ab}$. (The right hand
side is symmetric under the substitutions, $ \b \leftrightarrow \d $
or $ \a \leftrightarrow \g $.)  The corresponding quartic trace for
leptons is $ \calT _{1234}=1 $.  Identifying the leading pole term in
Eq.~(\ref{eqloexp}) to the pole term in the analogous field theory,
and using the low-energy limit with due account for the gauge factors
in Eq.~(\ref{eqtracecp}), one finds that the ratio of the string to
field theory gauge coupling constants of the non-Abelian gauge factors
should be set as, $\eta = \sqrt {2} $.

\subsection{Numerical results for contact interactions}
\label{subsect24} 

We start off by stating the main simplifications made in our numerical
study. We choose to set all the radii parameters, $ r_1 ^I $, equal to
a common radius parameter, denoted by $r = 1/ M_c$.  The relation
between the string theory parameters, the wrapped three-cycles volume
and gauge coupling constant parameters, $ g_s, \ m_s ,\ \call_A , \ g
_A , \ [A=a, b, c, d] $ simplifies then to \bea && m_s r = ({4\pi K_A
g_s \over g_A ^2 \vert \call_A \vert }) ^{1\over 3 } , \ [\call_A
=\prod _I {\vert L^I_A \vert \over r } = \prod _I (n_a ^{I2} + \tilde
m_a ^{I2} \chi ^{I2} ) ^\ud , \ \chi ^{I} = { r_2^I \over r } ] \eea
where $ K_{a,c,d} =1,\ K_b =2$.
From the explicit formulas for the three-cycles volumes, we infer the
relations between the branes gauge coupling constants \bea && \call_a
= \call_d= [ (\rho ^{-2} + ( 3 \rho \b _1 \chi _2 ) ^2 ) (\rho ^{-2} +
( 3 \rho \b _2 \chi _3 ) ^2 ) ] ^ \ud , \ \call_b = {\chi _1\chi _3
\over \b _1} , \ \call_c = {\chi _1 \chi _2 \over \b _2} \cr &&
\Longrightarrow \ g_a = g_d , \ {g_b^2 /2 \over g_c^2 } = {\call _c
\over \call _b } = {\b _1 \chi _2 \over \b _2 \chi _3} . \eea In the
case with $\caln =1 $ supersymmetry, $\b _2 \chi _3 = \b _1 \chi _2$,
one has, $ \call _c = \call _b $ and $ 2 g_b ^{-2} = g_c^{-2}$.
Specializing momentarily to this supersymmetric case and using the
proportionality relation between the string and field theory gauge
coupling constants discussed after Eq.~(\ref{eqstramp1}), we can
express the Standard Model gauge coupling constants $ g_3, \ g_2, \
g_1 $ along with the electric charge and weak angle parameters, $ e $
and $\sin \t _W$, by the formulas valid at the string mass scale \bea
&& \eta g_3 = g_a,\ \eta g_2 = g_b, \ ( \eta g_1 ) ^{-2}
= {1 \over 6} g_a ^{-2} + {1 \over 2} g_c ^{-2} + {1 \over 2} g_d
^{-2} = {2 \over 3} g_a ^{-2} + g_b ^{-2} , \cr && \ \Longrightarrow \
( \eta e ) ^{-2} = ( \eta g_1) ^{-2} + ( \eta g_2) ^{-2} = {2 \over 3}
g_a ^{-2} + 2 g_b ^{-2}, \ \sin ^2 \t _W \equiv {g_2 ^{-2} \over e
^{-2} } = {3 g_b ^{-2} \over 2 g_a ^{-2} + 6 g_b ^{-2} } = {1 \over 2
( 1 + g_b ^{2} / (3 g_a ^{2} ) ) } . \label{eqrelccs} \eea

Although the family of models under consideration has a parameter
space of restricted size, to study the model dependence of predictions
we found it convenient to introduce a reference set of natural values
for the geometric parameters and consider small excursions in which
the parameters are varied one by one.  We define our reference set of
parameters as, $\rho =1, \e = 1, \tilde \e =1, \b _1 =1, \b _2 =1$.

The interbrane angle parameters in the three complex planes, $\t ^I
_{\mu \nu } = \t ^I _{\nu } - \t ^I _{\mu },\ [\pi \t ^I _{\mu } =
\arctan ( m_\mu ^I \chi ^I /n_\mu ^I) ] $ are grouped into two
distinct sets with the entries in each set being equal up to
permutations of the planes.  For the reference set of parameters with
$\chi ^I =1$, we find $\t ^I ( q ) =\t ^I (e ) =( 0.50, 0.60, 0.89 ),$
and $ \t ^I ( u^c ) = \t ^I ( \nu ^c ) \simeq \t ^I ( d^c ) = \t ^I (
e^c ) = (0.50, 0.89 , 0.60).  $ Let us also quote, in reference to the
discussion near Eq.(\ref{eqformf}), the numerical values assumed by
the form factor parameter, $\ln \d ^I =( 2.77,\ 2.97,\ 10.1)$, for the
above quoted values of the interbrane angles for $\t ^I ( q ) $.
Varying the tori complex structure parameters inside the range, $ \chi
^I \in [\ud , 2]$, or changing from orthogonal to tilted tori, causes
insignificant changes in the angles.  For instance, the choice $ \b
_1= \b _2= \ud $ yields $\t ^I( q ) =(0.5, 0.687, 0.812)$.

With the free parameters consisting of $m_s $ and $m_s r = m_s / M_c$,
the string coupling constant is fixed in terms of the gauge coupling
constants by, $ g_s = {m_s ^3 \vert L_A \vert g_A ^2 \over 4 \pi K_A
}$.  For the string theory not to be driven to strong coupling in the
decompactification limit at fixed $m_s$, the condition $g_s \propto
(m_s r ) ^3 < 1$ restricts the radius parameter to $m_s r = O(1)$.
More precisely, setting tentatively in $ m_s r = ({4 \pi K_A g_s \over
g_A ^2 \vert \call _A \vert }) ^{1\over 3} ,\ [A=a, b] $ the Standard
Model gauge coupling constants to their observed values at the $Z$
boson mass scale, $ g_1^2 (m_Z)= 0.127 ,\ g_2^2(m_Z)=0.425,\ g_3^2
(m_Z)= 1.44 $, we can express the conditions that the string theory is
weakly coupled by the numerical results evaluated for the reference
set of parameters, $ m_s r \simeq [0.95 ,\ 3.9] g_s ^{1/3} /\eta
^{2/3} $.

The weak angle depends sensitively on the geometric parameters.  While
setting $ g_a = g_b$ reproduces the favored value, $ \sin ^2 \t _W =
{3\over 8} $, using the reference set of parameters with $\chi ^I=1$,
would yield instead, $ {g_a ^2 / (g_b ^2 /2) } = \call _b / \call _a
\simeq 1/ 10 \ \Longrightarrow \ \sin ^2 \t _W \simeq 3/46 $.
However, as verified from the explicit formula, $\sin ^2 \t _W = 1/[2
(1+ 2 \b _1 (1 + 9 \b_1 ^2 \chi _2 ^2 ) / (3 \chi _1 \chi _3 ) ) ] $,
one can always fit the weak angle by adjusting the geometric
parameters, for instance, by setting, $ \b _1 = 1/2 $ with all $\chi
^I $ equal. The renormalization group analysis of the gauge coupling
constants for the present model is known to be consistent with grand
unification with the parameters adjusted at~\cite{blumstieber03} $m_s
\simeq M_X \simeq 10^{16 } $ GeV for $m_s/M_c \simeq (2 \ - \ 5)$.  On
the other hand, the qualitative study of the coupling constant
unification for the class of minimal supersymmetric standard models
selected by sampling over vacuum solutions of intersecting branes on
orientifolds~\cite{anastaso06} indicates that the maximal allowed
string scale may vary over a wide interval, provided one regards the
branes gauge coupling constants as free parameters.  In order to
justify the TeV string scale scenario of interest to us, one must
invoke string threshold corrections which produce an accelerated power
law running between the compactification and string mass
scales~\cite{dienes98}.  Although the small extent of the admissible
interval for the momentum scale, $ Q \in [ M_c, m_s ]$, could render
this option problematic, some freedom is still left in choosing the
geometric parameters.  To verify this statement we have pursued a
qualitative study of the relation between the one-loop order running
gauge coupling constants holding for the present model, \bea && (4 \pi
)^2 \bigg [ {2 \over 3 g^2 _3 (m_Z ) } + { 2\sin ^2 \t _W (m_Z )
-1\over e^2 (m_Z) }\bigg ] - {44\over 3} \log {m_Z^2 \over m_s ^{2} }
\simeq -{\pi \hat b } ({m_s \over M_c })^2, \eea using similar inputs
as those discussed in our previous work~\cite{chem06}. The power law
running term on the right hand side is controlled by a linear
combination of slope parameters denoted by $\hat b $. We have checked
that with the assigned string mass scale, $ m_s \simeq 1 $ TeV, we can
satisfy the above relation by using the indicative values, $ \hat b
\sim 20 , \ {m_s \over M_c } \sim 5.$
 
\begin{table}
\begin{center}
\caption{\it Chirality conserving contact interactions of two quark
and/or lepton pairs of fixed chiralities.  In Column $1$, we specify
the four fermions configuration; in Column $2$, the target space
polygon $ DABC$ which realizes the world sheet embedding; in Column
$3$, the interbrane angles $\t , \ \t '$ associated to the two pairs
of conjugate fermions $ f, \ f' $; in Column $4$, the quartic traces
over the gauge matrices including the direct and reverse permutations,
$ \calT _{1234} = T_{1234} + T _{4321} $, multiplied by the Dirac
spinor matrix element, $ \cals _{1234} $, using the conventional
ordering for the flavor and color indices $ i, j, \cdots $ and $ \a ,
\b , \cdots $ of the incoming fermions, $ f_{1, i, \a } (-\t ) f_{2,
j, \b } (\t ) f_{3, k, \g } (-\t ') f_{4, l, \d } (\t ') $, with the
dependence on color indices defined by, $ T_ {\a \b \g \d } = \ud (\d
_{ \a \b } \d _{ \g \d } + \d _{ \a \d } \d _{ \g \b } ) $; and in
Column $5$, the numerical predictions in Cremades et
al.,~\cite{CIM,CIM2} model for the flavor diagonal coefficients $ g
^{HH'} _{ii,ii} \equiv ( m_s ^2 K_A / \L _{HH'} ^{ff'2} g_A ^2 ) $ in
Eq.~(\ref{eqcoefg}) at the three values of the compactification scale
parameter, $ m_s r = 1, \ 2,\ 3 $. }
\begin{tabular}{|cccc|c|} \hline &&&& \\ Fermions $ f^2 f^{'2}$ &
 (DABC) & $ \t , \ \t '$ & $ { 2 \pi g^2 _\mu \over K_\mu } \calT
_{1234} $ $\cals _{1234} $ & $ g_{ff'}^{HH'} (L_\mu ) $ \\ &&&& \\
\hline \hline &&&& \\ $ (e_L ) ^2 (e_L ) ^2 $ & $ (bdbd) $ & $ \t
_{db} ,\ \t _{db} $ & $ { 2 \pi g^2 _ d \over K_d } (\bar e_{1L} \g ^
\mu e_{2L} ) (\bar e_{3L} \g _ \mu e_{4L} ) $ & $ + 0.049 -0.270
-0.952 $ \\ &&&& \\ $ (e_L^c )^2 (e_L^c )^2 $ & $ (dc'dc') $ & $ \t
_{c'd} , \ \t _{c'd} $ & $ { 2 \pi g^2 _ c \over K_c } (\bar e_{2R} \g
^ \mu e_{1R} ) (\bar e_{4R} \g _ \mu e_{3R} ) $ & $ + 0.049 -0.270
-0.974 $ \\ &&&& \\ $ (e_L )^2 (e_L^c )^2 $ & $ (dc'db) $ & $ \t
_{c'd} , \ \t _{bd} $ & $ { 2 \pi g^2 _ c \over K_c } \ud (\bar e_{2R}
\g ^ \mu e_{1R} ) (\bar e_{4L} \g _ \mu e_{3L} ) $ & $ -0.405 -1.92
-5.74 $ \\ \hline \hline &&&& \\ $ (q_L )^2 (q_L )^2 $ & $ (baba) $ &
$ \t _{ab} , \ \t _{ab} $ & $ { 2 \pi g^2 _ a \over K_a } T_{\a\b \g
\d } (\bar q_{1L} \g ^ \mu q_{2L} ) (\bar q_{3L} \g _ \mu q_{4L} ) $ &
$ + 0.049 -0.270 -0.952 $ \\ &&&& \\ $ (u_L^c )^2 (u_L^c )^2 $ & $
(acac) $ & $ \t _{ca} , \ \t _{ca} $ & $ { 2 \pi g^2 _c \over K_c }
T_{\a\b \g \d }(\bar u_{2R} \g ^ \mu u_{1R} ) (\bar u_{4R} \g _ \mu
u_{3R} ) $ & $ 0.0422 -0.286 -1.015 $ \\ &&&& \\ $ (d_L^c )^2 (d_L^c
)^2 $ & $ (ac'ac') $ & $ \t _{c'a} , \ \t _{c'a} $ & $ { 2 \pi g^2 _ c
\over K_c} T_{\a\b \g \d } (\bar d_{2R} \g ^ \mu d_{1R} ) (\bar d_{4R}
\g _ \mu d_{3R} ) $ & $ + 0.049 -0.270 -0.974 $ \\ &&&& \\ $ (q_L )^2
(u_L^c )^2 $ & $ (abac) $ & $ \t _{ba} , \ \t _{ca} $ & $ { 2 \pi g^2
_ b \over K_b } \ud T_{\a\b \g \d } (\bar q_{2L} \g ^ \mu q_{1L} )
(\bar u_{4R} \g _ \mu u_{3R} ) $ & $ -0.388 -1.85 -5.54 $ \\ &&&& \\ $
(q_L )^2 (d_L^c )^2 $ & $ (abac') $ & $ \t _{ab} , \ \t _{c'a} $ & $ {
2 \pi g^2 _ b \over K_b } \ud \d _{\a\b \g \d } (\bar q_{2L} \g ^ \mu
q_{1L} ) (\bar d_{4R} \g _ \mu d_{3R} ) $ & $ -0.405 -1.92 -5.74 $ \\
\hline \hline &&&& \\ $ (q_L )^2 (e_L )^2 $ & $ (babd) $ & $ \t _{ab}
, \ \t _{db} $ & $ { 2 \pi g^2 _a \over K_a } \ud \d _{\a \b } (\bar
q_{1L} \g ^ \mu q_{2L} ) (\bar e_{3L} \g _ \mu e_{4L} ) $ & $ 0.049
-0.270 -0.974 $ \\ &&&& \\ $ (u_L^c )^2 e_L^c )^2 $ & $ (cacd') $ & $
\t _{ac} , \ \t _{d'c} $ & ${ 2 \pi g^2 _a \over K_a } \ud \d _{\a \b
} (\bar u_{1R} \g ^ \mu u_{2R} ) (\bar e_{4R} \g _ \mu e_{3R} ) $ & $
-0.165 -1.43 -4.15 $ \\ &&&& \\ $ (d_L^c )^2 (e_L ^c )^2 $ & $
(c'ac'd) $ & $ \t _{ac'} , \ \t _{dc'} $ & $ { 2 \pi g^2_a \over K_a }
\ud T_{\a \b } (\bar d_{1R} \g ^ \mu d_{2R} ) (\bar e_{3R} \g _ \mu
e_{4R} ) $ & $ -0.0157 -0.835 -2.22$ \\ \hline \hline
\end{tabular} \label{tabmod}
\end{center}
\end{table}

We now turn to the predictions for the chirality conserving contact
interactions of four quarks and/or leptons.  In view of the
uncertainties on the renormalization group scale evolution of the
gauge coupling constants and the coefficient functions, we have chosen
to express $g_s $ by selecting the electroweak gauge coupling
constant, $ g_b =\eta g_2 $, in the defining formula $ g_s = {\vert
\call _b \vert \eta ^2 g_2 ^2 (m_Z) \over 8 \pi }$, with the
proportionality factor set at $\eta = 1$.

The information on the branes configurations and on the structure of
the associated coefficient functions is displayed in
Table~\ref{tabmod} for Cremades et al.,~\cite{CIM,CIM2} model.  The
volume of cycles may vary substantially from one brane stack to the
other, so it is important to keep track of the data assigned to the
brane configurations $ DABC$ which affect the normalization factor of
the local operators.  Since we could not find any analytic
approximation that yields reliable estimates for the coefficients $
g^{HH'} _{ij,kl} $ in the appropriate range of $ m_s r$ values, we
have numerically evaluated Eq.~(\ref{eqcoefg}) by following the same
procedure described in our previous work~\cite{chem06}.  The
contributions from the localization and classical partition function
factors in the $x$-integral are evaluated by numerical quadrature
after removing the massless and momentum mode contributions by
subtracting the leading terms near the end points $ x\to 0 $ and $
x\to 1$, according to the prescription described schematically by
Eq.~(\ref{eqsubtr}).  The series summations over the world sheet
instantons must be carried out to large enough orders, $\max (p_A)
\simeq 6 - 10$ and the $x$-integral must be evaluated with care.  (We
have made use of the Mathematica package.)

The numerical values of the flavor diagonal ($\D F=0$) coefficients, $
g ^{HH'} _{ii,ii} $, obtained with the reference set of parameters at
the three values of the effective radius parameter, $ m_s r=1,\ 2, \ 3
$, are displayed in the last three columns of Table~\ref{tabmod}.
These results all refer to the $\D F=0$ configurations with coincident
intersection points, $\e _A =\e_B =(0,0,0).$ The presence of strong
cancellations from competing terms appears clearly from the fact that
the coefficients change sign with variable $ m_s r$.  The results for
the various flavor and chirality configurations are seen to cluster
around two group of values associated with the pure and mixed
chirality configurations, $ f_{L,R}^4$ and $f_L^2 f_R^2$, which also
correspond to the cases with equal and unequal angles. The
coefficients in the unequal angles group, $ e_L^2 e_R ^2 ,\ q_L ^2
d_R^2, \ u_R^2 e_R^2 $, are roughly equal and separated by a gap of
about a factor $5\ - \ 10$ from the coefficients in the equal angles
group, $ q_L^4,\ e_L^4 ,\ u_R^4 ,\ d_R ^4 , \ e_R ^4,\ q_L^2 e_L^2 ,\
d_R^2 e_R^2 $.  Since the exchange of massive vector and axial vector
bosons contribute pure and mixed chirality amplitudes of same and
opposite signs, respectively, we infer from the comparison of the
coefficients with same and opposite chiralities that the string
excitations are akin to linear combinations of vector and axial vector
modes. The flavor diagonal coefficients all feature the power law
growth with the radius parameter, $ \vert g _{ii,ii} ^{HH'} \vert
\propto ( m_s r ) ^ {5/2} $.  It is worth noting that the approximate
representation in Eq.~(\ref{eqloexp}), subsuming the contributions
near the boundaries of the $x$-integral, would not reproduce the
observed power law dependence of the coefficients on $ m_s r$.

The observed regularities in the coefficients are partly accounted for
by the symmetric character of the brane setup in the model at hand.
Since the branes $ a $ and $d $ are always parallel in all three
planes, equal interactions are found for $ q _L^4 $ and $ e _L^4 $ and
for $ d _R^4 $ and $ e _R^4 $. Numerically close values are also found
for, $ u^{c4} _L ,\ d^ {c4} _L $ and $ e^{c4} _L$, due to the fact
that the brane angles in the various configurations are equal up to
permutations of the complex planes.  The mixed chirality coefficients
are larger because they involve brane configurations with two sets of
unequal angles.

\begin{center} \begin{figure} 
\epsfxsize=5 in \epsfysize= 4 in \epsffile{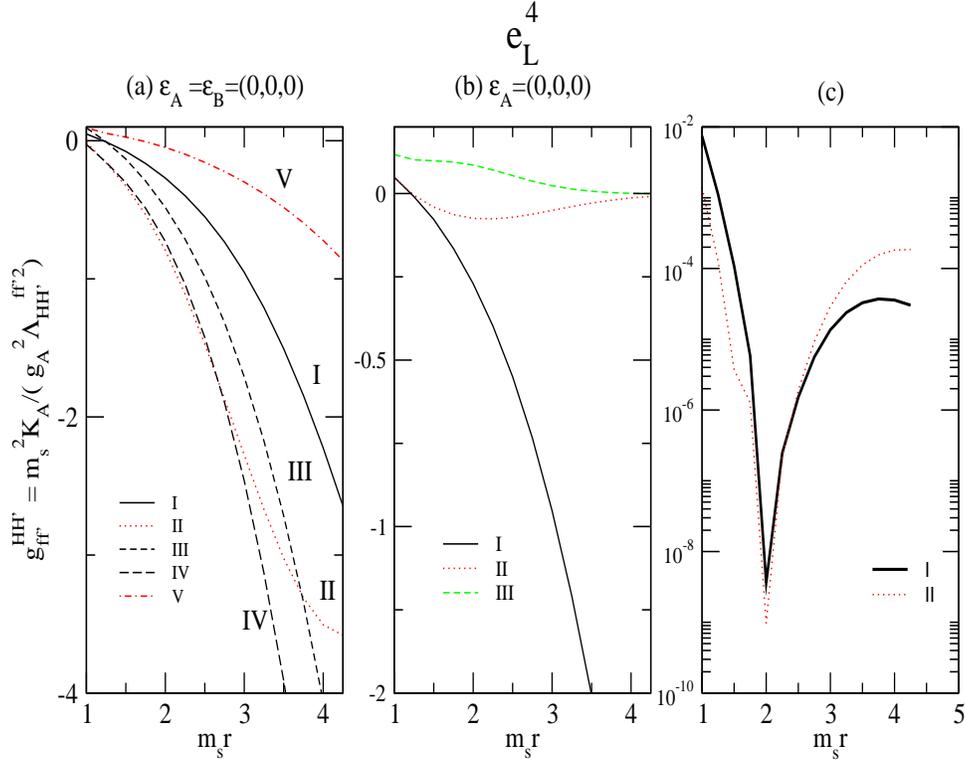}
\vskip 2 cm
\caption{\it The contact interaction coefficient $ g ^{HH'} _{ij,kl} =
m_s ^2 K_\mu / (\L ^ {ff'2}_{HH'} g_\mu ^2)$ for the operator $ e_L^4
$ predicted in Cremades et al.,~\cite{CIM,CIM2} model is plotted as a
function of $ m_s r$.  Same results hold for the operator $ q_L^4 $
and numerically close results hold for the operators $ u_R^4 ,\ d_R^4
$ and $ e_R^4$.  The reference set of parameters specifying the model
is defined by, $ \rho =1,\ \e =1, \tilde \e =1, \ \chi ^I = 1,\ \b_1=
1,\ \b_2= 1 ,\ \e _A =(0,0,0),\ \e _B =(0,0,0)$ (see the text just
below Eq.~(\ref{eqrelccs})).  In panel $(a)$, the five curves $ I, II,
III, IV, V$ are obtained by starting from the reference set $(I) $ and
performing in succession the following variations: $ \chi ^I = 2 \
(II) $, $ \b_2= 1/2 \ (III) $, $ \b_1= \b_2= 1/2 \ (IV) $ and $ \chi
^I =\ud \ (V) $.  In panel $(b)$, the three curves $ I, II, III $ are
obtained by starting from the reference set of parameters $(I) $ and
including the changes, $ \e _B =(0,{1\over 3} ,0) \ (II) $ and $ \e _B
=(0,{2\over 3} ,0) \ (III) $.  In panel $(c)$, the two curves $ I, II
$ refer to the reference set of parameters with the choices, $\e _A=
\e _B =(0,{1\over 3} ,0) \ (I) $ and $\e _A= \e _B =(0,{2\over 3} ,0)
\ (II) $.The presence of cusps in the curves of panel $(c)$ is due to
the use of semi-logarithmic plots for the absolute values of the
coefficients.}
\label{fig1} \end{figure} \end{center}
\begin{center} \begin{figure} 
\epsfxsize=5 in \epsfysize=4 in \epsffile{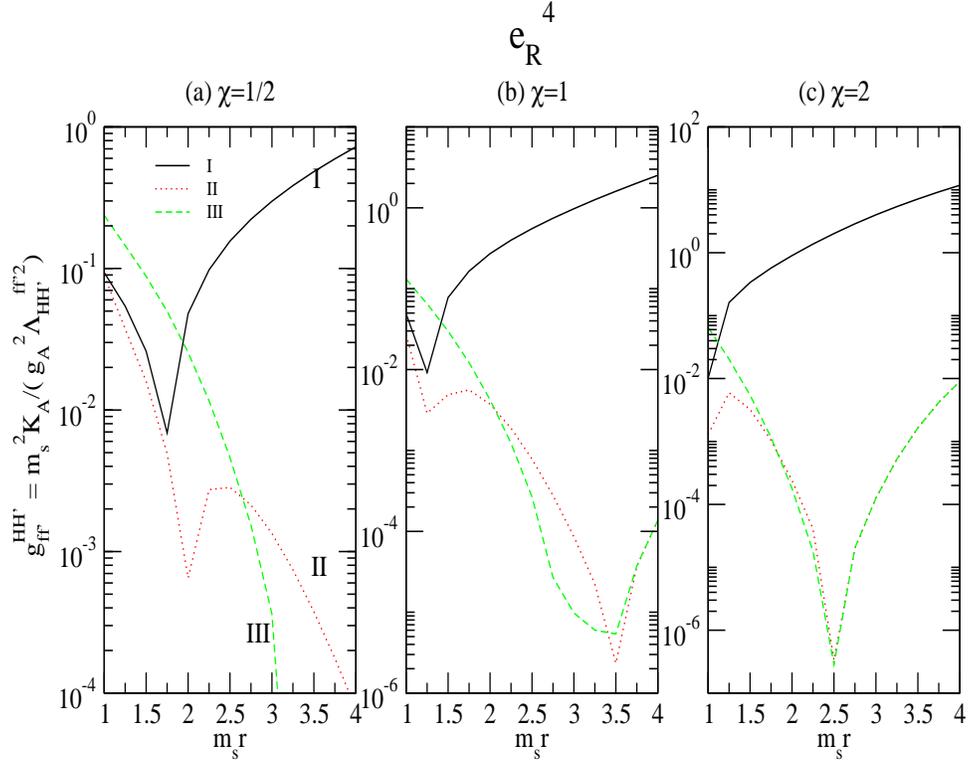}
\vskip 2 cm \caption{\it The contact interaction coefficient $ g
^{HH'} _{ij,kl} $ for the operator $ (e_L ^{c})^4 \sim e_R^4 $ is
plotted as a function of $ m_s r$.  We use same conventions as in
Fig.~(\ref{fig1}).  Same results hold for the operator $(d_L ^{c})^4
\sim d_R^4 $ and numerically close results hold for the operator $(u_L
^{c})^4 \sim u_R^4 $.  The three panels $ (a),\ (b), \ (c) $ are
associated to $ \chi ^I = \ud, \ 1,\ 2 $. In each panel, we display
three curves obtained by starting from the reference set of parameters
$(I) $ and changing the longitudinal distance parameter to $ \e _B
=(0,{1\over 3} ,0) \ (II) $ and $ \e _B =(0,{2\over 3} ,0) \ (III) $.
The presence of cusps in certain curves is due to the use of
semi-logarithmic plots for the absolute values of the coefficients.}
\label{fig2} \end{figure} \end{center}
\begin{center} \begin{figure} 
\epsfxsize=5 in \epsfysize=4 in \epsffile{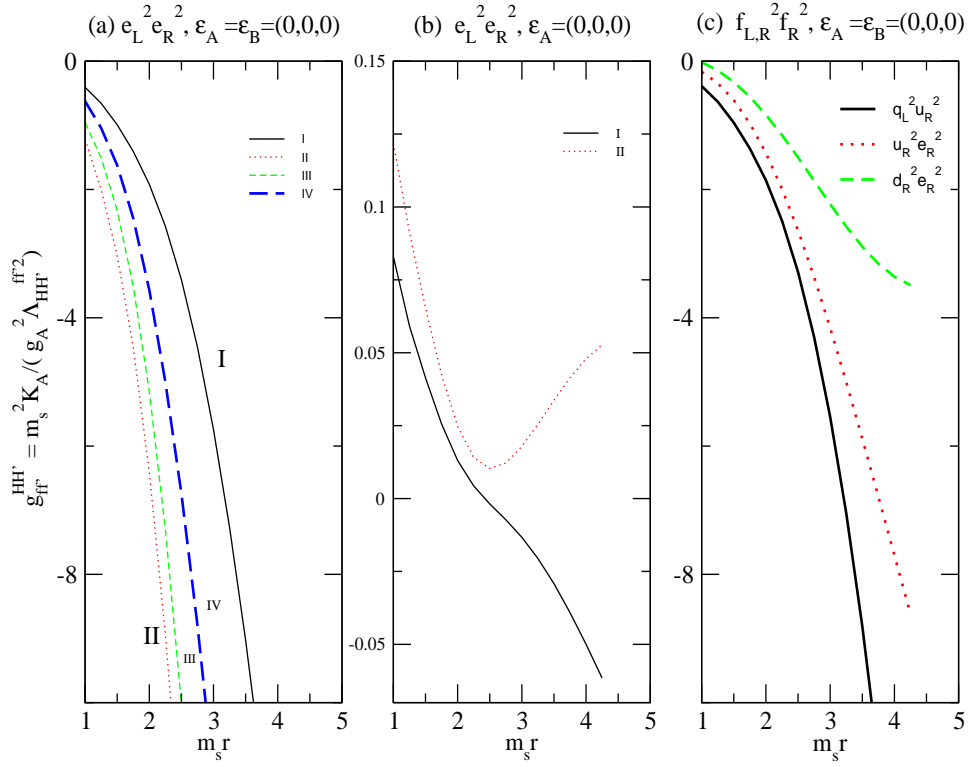}
\vskip 2 cm \caption{\it The contact interaction coefficients $ g
^{HH'} _{ij,kl} = m_s ^2 K_\mu / (\L ^ {ff'2}_{HH'} g_\mu ^2)$ are
plotted as a function of $ m_s r$ for various configurations of the
quarks and leptons. We use same conventions as in Fig.~(\ref{fig1}).
In the panel $(a)$ referring to the operator $ e_L^2 (e_L ^{c})^2 \sim
e_L^2 e_R^2 $, we display four curves obtained by starting from the
reference set of parameters $(I) $, and independently changing the
single parameters, $ \chi ^I = 2 \ (II) $, $ \b_2= 1/2 \ (III) $, $
\b_1= \b_2= 1/2 \ (IV) $.  In the panel $(b)$ referring to the
operator $ e_L^2 e_R^2 $, we display two curves obtained for the
reference set of parameters with the values of the longitudinal
distance parameter, $ \e _B =(0,{1\over 3} ,0) \ (I) $ and $ \e _B
=(0,{2\over 3} ,0) \ (II) $. In panel $(c)$, we display the
coefficients for the quark and quark-lepton operators, $ q_L^2 u_R^2
,\ u_R^2 e_R^2 ,\ d_R ^2 e_R ^2 $, obtained with the reference set of
parameters.}
\label{fig3} \end{figure} \end{center}

For a clearer assessment of the dependence on $ m_s r$, we display in
Figures~\ref{fig1}, \ref{fig2} and \ref{fig3} plots of the
coefficients $ g _{ij, kl} ^{HH'} $.  We study here the sensitivity on
the geometric parameters by varying these one by one with respect to
the reference set.  Note that the longitudinal distances $\e _A,\ \e
_B$ are associated with flavor change while the transverse distances
are associated with gauge symmetry breaking. We group the
configurations into three classes corresponding to the flavor change:
$\D F= 0 :\ \e _A =\e _B =0 ;\ \D F= 1 :\ \e _A =0,\ \e _B \ne 0 ;$
and $ \D F= 2 :\ \e _A \ne 0 ,\ \e _B \ne 0 $.  Figures~\ref{fig1} and
\ref{fig2} refer to the unmixed chirality configurations with equal
brane angles and Figure~\ref{fig3} to the mixed chirality
configurations with unequal brane angles. The presence of cusp
discontinuities in certain curves is due to our use of logarithmic
plots for the absolute values of the coefficients aimed at
representing quantitatively the size of the suppression.

We see on panel $(a)$ of Fig.~\ref{fig1} that the predictions are
spread by an approximate factor $2 \ - \ 3$ for reasonably restricted
variations of the shape parameters.  Using tilted 2-d tori, or
increasing the complex structure parameter $ \chi ^I $, results in
enhanced coefficients, while decreasing $ \chi ^I $ results in reduced
coefficients.  The $ \D F =1$ coefficients with finite $ \e _B $ in
panel $(b)$ are suppressed by order $10^{-1}$ while the $ \D F =2$
coefficients with finite $ \e _A $ and $ \e _B $ in panel $(c)$ are
suppressed by factors of order $ 10 ^ {-2} - 10 ^ {-4} .$ The specific
dependence on $m_s r$ is a result of the tension between the power
growth from the overall normalization factor $ g_s \propto (m_s r ) ^3
$ and the exponential suppression from $ Z_{cl}$.  That the
suppression effect is controlled by the classical factor is clear from
the fact that the coefficients have comparable values near $ m_s r
=1$.

The plots in Fig.~\ref{fig2} again confirm that $ \chi ^I < 1 $ and $
\chi ^I >1 $ lead to reduced and enhanced coefficients.  The nearly
one order of magnitude suppression of the $\D F=1 $ coefficients is
independent of $\chi ^I $.  That the suppression is weaker than
expected is explained by the specific feature in the present model
that only a single component of the vectors $ \e _B^I $ are finite.
The comparable predictions found for $\e _B = (0,1/3,0) $ and $
(0,2/3,0)$ are explained by the torus lattice periodicity.  The
cancellation effects from the oscillating factors $ e ^{2i\pi \e L} $
explain both the change of sign from positive to negative coefficients
and the smooth variation with $ m_s r$.

The plots in Fig.~\ref{fig3} show that the coefficients in the unequal
angle cases are systematically larger than those found for equal
angles.  The dependence of the wrapped cycle volumes on the torus
shape parameters spreads the coefficients by a factor of $ 2 \ -\ 3 $.

Two general features of the predictions are the rapid power law
increase with $ m_sr$ of the flavor conserving $\D F=0$ coefficients
and the hierarchies of order $ 10^{-1} $ and $ 10^{-4}$ separating
these from the flavor changing $\D F=1 $ and $\D F= 2$ coefficients
which vary more slowly over the allowed interval for $m_sr $.  While
the variation of the coefficients with $m_s r$ is not apparent on the
result in Eq.~(\ref{eqloexp}), obtained by restricting the
$x$-integral to the end point contributions, it appears possible to
use this approximate formula in order to explain the dependence on the
distance parameters, $\e _A,\ \e_B$.  Examination of the combined
contributions from the string momentum and winding modes to the
coefficients \bea && g ^{HH'} _{ff'} \simeq {\vert L_A\vert \over 4
\pi \vert L_B \vert } \sum _{p_B\ne 0 } \sum _{p_A \in Z } { \prod _I
\d _I ^{ - \vert \sin (\pi \t ^I) (p_A +\e _A ^I) L_A ^I \vert ^2 +
\vert p_B / L_B ^I \vert ^2 } e ^{ 2 \pi i p_B \e_B ^I} \over \sum _I
\vert \sin (\pi \t ^I) (p_A +\e _A ^I) L_A ^I \vert ^2 + \vert p_B /
L_B ^I \vert ^2 } , \eea shows that for small finite $ \e _A$ the
leading contribution to the ratio of $\D F=1 $ to $ \D F=0$ amplitudes
is of form $ e ^{- p_A \e _A \vert \call _A \vert (m_s r)^2 }$, while
the $ \D F=2$ amplitudes include the additional suppression from the
oscillating factors, $ e ^{2\pi i \e _B } $.

It is interesting to compare our predictions for the contact
interactions of four fermions with those obtained in the
$Dp/D(p+4)$-brane models~\cite{antbengier01}. (The formalism is
briefly reviewed in Appendix~\ref{sectappen1}.)  For the coupling of
four modes, $ \vert (3,7_I) \vert ^2 \vert (3,7_J) \vert ^2 $, the
comparison at fixed 4-d gauge coupling constant of our estimate, $ 2
\pi g _ {ff'}^{HH'} \cals _{1234} \calT _{1234} \approx 2 \pi (0.05 \
- \ 0.5) (m_sr )^2 \cals _{1234} \calT _{1234} $, with the result
found by Antoniadis et al.,~\cite{antbengier01} in the large radius
limit, $ \calT _{1234} [ 0.12 \ \calp _{1234} + 0.33\ \cals _{1234} ]
$, reveals an order of magnitude concordance.

Finally, we compare our predictions with the contributions from the
momentum modes evaluated by means of Eq.~(\ref{eqcontkk}) for $
n=3$. The resulting rough estimate, $ (g _ {ff'}^{HH'})_{KK}
\simeq {1 \over 4 \pi (n-2)} { S_n (m_s r )^n \vert \call _B \vert }
\simeq (m_s r )^3 \call_B $, indicates that the contributions from the
string momentum modes are significantly larger than those from winding
modes. We should remember, however, that the present estimate must be
regarded as an upper bound since it relies on the large $r$ limit and
ignores the form factor suppression.

\section{Indirect high energy collider tests}
\label{sect3}

We discuss in the present section the collider physics applications
based on the formalism presented in Subsection~\ref{subsect22} and in
Appendix~\ref{sectappen1} for the orientifold model of Cremades et
al.,~\cite{CIM,CIM2}.  Since the distinction between the mass and
gauge bases is not essential for these observables, all the results in
this section are obtained by setting the flavor mixing matrices to
unity, $ V _H ^f =1$.

\subsection{Bounds on contact interactions   mass scales}

It is important to distinguish the mass scale $\L $ associated to the
$\cddd =6 $ operators from the mass scale $M_H$ associated to the
$\cddd =8 $ operators in the 4-d effective Lagrange density quadratic
in the energy-momentum tensor~\cite{hewett99}, $ L_{EFF} = i{ 4 \l
\over M_H^4} T_{\mu \nu } T^{\mu \nu } $.  The analyses of available
high energy collider experimental data using field theories in extra
space dimensions are sensitive to values of these mass scales, $M _H =
1.5 $ TeV~\cite{hewett99}, $\L = 2 \ - \ 6 $ TeV~\cite{rizzo99} and
$\L = 1 \ - \ 8 $ TeV~\cite{antobequ99}.  In the single $Dp$-brane
models, the quantum gravity mass scale $M_H$ was found to be
parametrically larger than the string scale~\cite{cpp00}, ${ M_H \over
m_s } \simeq {2 ^{3/4} \over \pi \sqrt {g_s} } $, thus making the
detection of new physics effects harder.  To pursue the comparison
with the intersecting brane models, it is convenient to consider in
place of $ M_H$ the closely related gravitational mass scale $ M_{G} $
defined in the case with $n$ flat extra dimensions by~\cite{add99}, $
M_{G} ^{n+2} r ^ {n} = {M_P ^2 \over 4\pi } = (4\pi G_N)^{-1} $.  The
mass scale $ M_G$ is related to the fundamental string parameters of
single $D3_A$-brane models as~\cite{cpp00}, $ ( {M_G/ m_s} )^8 = {16
\pi / g_A^4 } .$ Repeating the same calculations for intersecting
$D6_A$-branes gives us the modified formula in the large radius limit
\bea && ({M_G\over m_s} )^8 = {16\pi \over g_A^4 (m_s r )^6 \vert m_A
^I- n_A ^I U^I \vert / U_2 ^I } .\eea The strong dependence on the
geometric parameters indicates the interesting possibility that $M_G$
may assume lower values in multiple brane models.

We now discuss the constraints on the string mass parameter, $m_s$,
inferred by comparing our predictions for the chirality conserving
contact interactions of $ \cddd =6$ with a subset of the available
experimental limits~\cite{pdg06}.  For the lepton and lepton-quark
configurations, $ e_L^4 ,\ e_L^2 q_L^2$ and the quark configuration,
$q_L^4 $, respectively, we evaluate the bounds on the string scale
parameter, $ m_s ^2 =g ^{LL}_{ii,ii} \L ^{ef 2} _{LL} (1+\d _{ef})
g_A^2 / K_A $, at fixed $ m_s r$, by setting the gauge coupling
constants, $ g_A ^2 /K_A $ at $ g_2^2/2 \simeq 0.213 $ and $g_3 ^2
\simeq 1.44$, respectively.  Using the numerical values of the
coefficients in Table~\ref{tabmod}, we obtain the following bounds on
$ m_s$ for the choice of three representative experimental limits on
the mass scales: \bea && \L ^\pm _{LL} ( ee ee) > [ 4.7 , \ 6.1 ] \
\Longrightarrow \ m_s > [0.69,\ 2.1, \ 3.9 ] \cr && \L ^\pm _{LL} ( ee
qq) > [ 23.3 , \ 12.5 ] \ \Longrightarrow \ m_s > [2.4 ,\ 3.0, \ 5.6]
\cr && \L ^\pm _{LL} ( qq qq) > [2.7]\ \Longrightarrow \ m_s > [ 0.72
,\ 1.7 , \ 3.2 ] ,\eea where all masses are expressed in TeV units and
the three entries refer to the values $ m_sr= [1, \ 2,\ 3 ] $.

We next consider the constraint from the enhanced supernova cooling
through the reaction producing right handed neutrino-antineutrino
pairs by quark-antiquark pairs, $ q + \bar q \to \nu ^c+ \bar \nu ^c
$, which is allowed as long as the contributions to the neutrino Dirac
mass are bounded by the supernova temperature, $ m_\nu \leq T _{SN}
\simeq 50 $ MeV.  The lower bound on the mass scale in the effective
Lagrangian, $ L _ {EFF}= {4 \pi \over \L ^{q \nu _R 2 }} (\bar q
\g_\mu \g_5 q ) (\bar \nu _R \g ^\mu \nu _R ) $, is found for the
SN1987A to cover the range~\cite{grifols98}, $ \L ^{q \nu _R} > ( 90 \
- \ 250 ) $ TeV.  For concreteness, we set our choice on the lower
bound, $ \L ^{q \nu _R} > 200 $ TeV.  Using the numerical predictions
in Table~\ref{tabmod} and assuming the approximate equalities between
the chirality basis amplitudes, $ q_L^2 \nu _R^2 \simeq q_L^2 u _R^2 $
and $ d_R^2 \nu _R^2 \simeq d_R^4$, we obtain the numerical estimate
for the coefficient, $ (g ^{LR} _{q _L\nu _R} +g ^{RR} _{q _R\nu _R} )
\simeq [0.66,\ 1.26,\ 2.14] $ for $ m_sr= [1, \ 2,\ 3]$. The resulting
bounds on the string mass scale read, $ m_s > [43,\ 82,\ 139 ]$ TeV.
For reference, we note that the comparison with the contribution from
the string momentum modes yields~\cite{abeleb03} the weaker bound, $
m_s \geq (5 \ - \ 10 ) $ TeV.

\subsection{Bhabha scattering  cross section}

Useful constraints on the new physics are set by the experimental data
at the high-energy colliders involving the two-body
processes~\cite{lep2rev} of Bhabha, M\"oller and photon pair
scattering and fermion-antifermion pair production.  The absence of
significant deviations from the Standard Model predictions has led the
statistical analyses of data to set exclusion limits on the free
parameters.  The global fits to the combined high-energy collider data
based on the single $D$-brane model, with the gauge factors treated as
free parameters, yields~\cite{cheung05} $m_s > (0.69 \ - \ 1.96) $
TeV.

We focus here on the Bhabha scattering differential cross section for
which high precision measurements along with higher order calculations
of the pertubation theory corrections are due in the future.
Experimental data has been collected by the LEP
collaborations~\cite{l399,aleph99,opal99}.  The studies based on
single $Dp$-brane models, describing the ratio of the string to
Standard Model differential cross sections by the approximate
formula~\cite{cpp00} \bea && R (\cos \t ) \equiv {(d \s / d \O ) \over
(d \s / d \O )_ {SM} } = \vert \cals (s,t) \vert ^2 , \cr && [\cals
(s,t) \equiv { \G (1-s) \G (1-t) \over \G (1-s-t) } \simeq 1 - {\pi ^2
st \over 6 m_s ^4 } +\cdots ] \label{eqsngl} \eea
yield by comparison with the experimental data at the center of mass
energy $ \sqrt s = 183 $ GeV the $ 95 \% $ confidence level exclusion
limit on the string mass scale, $ m_s > 410$ GeV.  Similar bounds are
found in the analysis~\cite{bouril99} including the experimental data
at $ \sqrt s = 188.7 $ GeV.  At the higher energy, $ \sqrt s = 1 $
TeV, the $ 95 \% $ confidence exclusion limit obtained under similar
conditions
should extend the sensitivity reach to~\cite{cpp00}, $ m_s > 3.1 $
TeV.


We now present our predictions for the Bhabha scattering differential
cross section evaluated with Cremades et al.,~\cite{CIM,CIM2} model by
adding the contributions from the contact interactions in
Eq.~(\ref{eqsubtr}) to the Standard Model amplitudes, using the
formalism detailed in Appendix~\ref{subsecapp2}.  In Fig.~\ref{fig4}
and Fig.~\ref{fig5}, we show plots of the ratio $R (\cos \t ) $ as a
function of the scattering angle variable, $ \cos \t $, for the center
of mass energies, $ \sqrt s =183$ GeV and $ \sqrt s = 500$ GeV,
respectively.  The selected set of values for $ m_s$ are different for
these two cases, as dictated by the fact that the string corrections
scale as $ s/m_s ^2$.

For a qualitative comparison with experimental data, we note that the
LEP data points for the ratio at $\sqrt s = 0.183 $ TeV are spread
over the interval of $\cos \t \in [-1, \ +1]$ inside the band limited
by the horizontal lines at, $R (\cos \t ) = (1.0\pm 0.4 )$. As for the
single $D$-brane model prediction~\cite{cpp00} in Eq.(\ref{eqsngl}),
this is represented by a nearly straight line which slopes from $ 1.06
$ to $ 1.0$ as $\cos \t $ increases from $ -1 $ to $ 1 $.  The
exchange of string Regge and winding modes are seen to give a small
reduction of $R (\cos \t ) $ near the forward scattering angles, $
\cos \t \sim 1$, gradually turning into a large enhancement near the
backward angles, $ \cos \t \sim -1$.  This implies a change from a
relative negative sign to a positive sign at some intermediate angle
in the interval $\cos \t \in [-1, \ +1]$.  The contributions grow
rapidly with increasing $m_s r $.  Requiring the predicted ratios in
Figs.~\ref{fig4} and~\ref{fig5} to remain bounded inside the interval
$R (\cos \t )\in [0.8, 1.2] $ for $\cos \t \in [-1, \ +1]$ imposes
lower bounds on the string scale which cover the ranges, $ m_s\geq
(0.5\ - \ 3.) $ TeV and $ m_s\geq (2.\ -\ 5.)$ TeV, respectively, for
the interval of values $ m_s r \in [1, 3] $.

We have also performed a more realistic calculation of the Bhabha
scattering differential cross section in which the total regularized
string amplitudes are obtained by subtracting by hand the pole terms
from exchange of massless and massive momentum modes for the $\g \ +
Z$ gauge bosons, while adding up the contributions from the physical
$\g \ + Z$ pole terms, using the prescription in Eq.~(\ref{eqsubtr}).
To ease the numerical calculations we have only evaluated the pure
chirality amplitudes $LL, \ RR$, while assuming the mixed chirality
amplitudes $ LR, \ RL$ to be proportional to these. The model
dependence on the ratio of pure to mixed chirality amplitudes thus
resides in the adjustable proportionality constant, $x = G _{e^c_L e_L
} / G _{e_L e_L } = G ^{RL} / G ^{LL} = G ^{LR} / G ^{LL} $, which we
have taken to vary inside the interval, $x = [\ud , \ 5].$

The ratio $R (\cos \t ) $ of the predicted differential cross section
to that of the Standard Model is plotted in Fig.~\ref{fig6} at the
center of mass energies, $ \sqrt s = 183 $ and $ 500 $ GeV (left and
right hand panels) for the two values of the string scale, $m_s = 1, \
2 $ TeV and $m_s = 2, \ 4 $ TeV, respectively.  The comparison of the
curves $I, \ II$, and similarly of the curves $ III, \ IV$, measures
the sensitivity of $ R(\cos \t )$ with respect to the string scale
$m_s$. On the other hand, the comparison of the curves $I, \ III$, and
similarly of the curves $ II, \ IV$, measures the sensitivity with
respect to the mixed chirality amplitudes.  The large spread of
predictions with variable $x$ and $m_s$ shows that Bhabha scattering
can usefully test the model dependence.  The results from the present
complete calculation agree qualitatively with those in
Figs.~\ref{fig4} and~\ref{fig5}, although the change of slope and
subsequent rise of the ratio with decreasing $\cos \t $ are generally
less steep.  We conclude that the representation of string amplitudes
by contact interactions is reliable for the considered incident
energies.


\section{Flavor changing neutral current processes}
\label{sect4}

\subsection{Direct and indirect flavor changing effects}
\label{subsect41}

The flavor changing neutral current observables are determined by the
non-diagonal elements of the mass basis coefficients of contact
interactions, $ \tilde G_{ij, kl} $. These receive direct flavor
changing contributions from the four point string amplitudes and
indirect contributions from the linear transformations linking the
gauge and mass bases of the fermions. Without further input
information on the flavor structure, it appears impossible to infer
quantitative constraints from a comparison with the flavor changing
observables.

A natural description of the fermions flavor quantum numbers is
provided by the basis labeled by the branes intersection points.  The
geometric constraints on string amplitudes, as stated by
Eq.~(\ref{eqgeomsel}), directly translate as selection rules on the
flavor amplitudes in this basis.  For Cremades et al.,~\cite{CIM,CIM2}
model, however, these rules turn out to be trivial ones, owing to the
fact that the intersection points for the fixed chirality modes lie at
finite distances apart only in single complex planes.  Since the
conditions involve at least one shift vector defined modulo $1$, no
zero entries are enforced either on the trilinear Yukawa interactions,
$\l ^f _{ij} f _i f^c_j H$, or on the four fermions interactions, $
G_{ij,kl} \bar f_i f_j \bar f_k f_l $.  This property is also
responsible for the separable structure of the Yukawa coupling
constants, $\l ^f _{ij} = a_i b_j$, implying that the mass matrices of
quarks and leptons all have unit rank.  Although the flavor
non-diagonal coefficients are generally finite, the $ \D F = 1, \ 2 $
operators associated with the configurations, $ i=j\ne k = l$ and $ i
\ne j \ne k $, are strongly suppressed with increasing $m_s r $ by the
classical partition function factor for longitudinal distance
parameters, $ \e _A ,\ \e _B $ of $O(1)$.  However, the fact that the
tree level string amplitudes depend on the relative distances between
intersection points, $ \e ^A _{ ij} $ and $\e ^B _{jk}$, introduces
certain restrictions on the flavor structure of the coefficients $ G
_{ij,kl} $.

We here focus on the two flavor changing neutral current observables
associated to the mass splitting of CP conjugate pairs of neutral
mesons made of quark-antiquark pairs, $ P=q_i \bar q _j,\ \bar P =
\bar q _i q_j$, and the three-body decays of leptons, $ l_j \to l_i +
l_k + \bar l_l .$ For simplicity, we assume that the Standard Model
contributions to these observables are negligible relative to those
from the contact interactions, so that we can directly use the
experimental limit to derive bounds on the string scale.  Convenient
formulas for the contributions to these observables from the chirality
conserving local operators have been obtained in~\cite{langplum00} for
models with extra $ U(1)$ gauge symmetries.  We follow closely the
formalism developed in the latter work, while accounting for the fact
that the dependence on the color quantum numbers is different in our
case.  The observables for the real and imaginary parts of the $P ^0 -
\bar P ^0 $ mass splitting, $ \D m_P \equiv - \Re (<P _0 \vert L ^{D S
=2} _{EFF} \vert \bar P ^0 > ) $ and $\e _P \equiv - \Im (<P _0 \vert
L ^{D S =2} _{EFF} \vert \bar P ^0 > ) / ( 2 \sqrt 2 m_P ) $, are
given in our notations by the explicit formulas \bea && \D m_P=
- 2 m_P F_P ^2 [ {1 \over 3} \Re ( \tilde G^{LL} _{ij,ij} + \tilde
  G^{RR} _{ij,ij} ) - \a '_{ij} \Re (\tilde G^{LR} _{ij,ij} ) ], \cr
  && \e _P = { m_P F_P ^2 \over \sqrt 2 \D m_P} [ {1 \over 3} \Im (
  \tilde G^{LL} _{ij,ij} + \tilde G^{RR} _{ij,ij} ) - \a '_{ij} \Im
  (\tilde G^{LR} _{ij,ij} ) ] , \eea

where $\a '_{ij} = {1 \over 3} + {2 \over 3} ( {m_P \over m _{q_i} +
m_{q_j} } ) ^2 $.  Note that the definition for the indirect CP
violation observable applies specifically to the $ K-\bar K $ system,
$\e _P = \e _K$.  The quarks flavor indices $ i, \ j $ are set in
accordance with the conventional assignments for the neutral mesons, $
K ^0 (\bar s d), \ B ^0 (\bar b d), \ B_s ^0 (\bar b s), \ D^0 (\bar u
c ), $ with $ F_P$ denoting the mesons two-body leptonic decay
coupling constants.  We have evaluated the hadronic matrix elements of
the four fermion local operators for the pseudoscalar mesons to vacuum
transition by making use of the vacuum insertion approximation. The
bilinear axial current hadronic matrix elements are determined through
the PCAC hypothesis in terms of the measured parameters $ F_P$.  Using
the conventional definitions for the matrix elements of quark bilinear
current operators \bea && <0\vert \bar s _{\a L} \g ^\mu d _{\b L}
\vert K^0(p)> = - <\bar K^0(p) \vert \bar s _{\a L} \g ^\mu d _{\b L}
\vert 0 > = i {F_K \over 6 \sqrt {2 m_K} } p^\mu \d _{a \b } ,\cr &&
<0 \vert \bar s _{\a } \g _5 d _{\b } \vert K^0(p)> = <\bar K^0(p)
\vert \bar s _{\a } \g _5 d _{\b } \vert 0 > = - i {F_K m_K^2 \over 3
\sqrt {2 m_K} (m_s + m_d ) } \d _{a \b } ,\eea one can write the
matrix elements of the relevant quadratic operators as \bea && Q^{LL}
_{ \a \b \g \d } \equiv < \bar K ^0 \vert (\bar s_{\a L } \g ^ \mu
d_{\b L } ) (\bar s_{\g L } \g ^ \mu d_{\d L } ) \vert K ^0 > = { F_K
^2 m_K \over 36 } (\d _{ \a \b }\d _{ \g \d } + \d _{ \a \d }\d _{ \g
\b } ) , \cr && Q^{LR} _{ \a \b \g \d } \equiv <\bar K ^0 \vert (\bar
s_{\a L } \g ^ \mu d_{\b L } ) (\bar s_{\g R } \g ^ \mu d_{\d R } )
\vert K ^0 > = - {F_K ^2 m_K \over 36 } [ \d _{ \a \b }\d _{ \g \d } +
2 ({m_K \over m _{q_i} + m _{q_j} } ) ^2 \d _{ \a \d } \d _{ \g \b } ]
, \cr && \ \Longrightarrow \ Q^{LL} _{ \a \b \g \d } T_{\a \b \g \d }
= {F_K^2 m_K \over 3} ,\ Q^{LR} _{ \a \b \g \d } T_{\a \b \g \d } = -
{F_K^2 m_K \over 3} [ 1 + 2 ( {m_K \over m_s + m_d } ) ^2 ] , \eea
where the saturation of color indices displayed in the last entry
above uses the tensor, $ T_{\a \b \g \d } = \ud (\d _{ \a \b } \d _{
\g \d } + \d _{ \a \d } \d _{ \g \b } ) $.  The factor $\a '_{ij} $
accompanying the coefficient $\tilde G ^{LR } _ {ij,ij} $, differs
from that quoted in~\cite{langplum00}, $ \a _{ij} = \ud + {1 \over 3}
({m_P \over m _{q_i} + m _{q_j} }) ^2 $, which refers to the
dependence on color indices involving the diagonal tensor, $ \d_{\a \b
} \d _{\g \d } $. Similar formulas hold for the $B $ and $ D $ mesons.

The contributions from contact interactions to the lepton number
violating three-body decay rates of charged leptons are given
by~\cite{langplum00} \bea && \G ( e_j \to e_i + e_k + \bar e_l ) = { m
_{e_j} ^5 \over 384 \pi ^3 } [ ( \vert \tilde G ^{LL}_{ji, kl} \vert
^2 + \vert \tilde G ^{LL}_{ji, kl} \vert ^2 + \vert \tilde G
^{LR}_{ji, kl} \vert ^2 + \vert \tilde G ^{LR}_{ji, kl} \vert ^2 ) + (
L \leftrightarrow R ) ] .  \eea The partial rates for the pair of
decay reactions, $\mu ^-\to e^- + e^+ + e^- $ and $\tau ^-\to e^- +
e^+ + e^- ,$ with $ j=2,\ 3$ and $ i=k=l=1$, are described by the
simplified formula \bea && \G ( e_j \to e_i + e_i + \bar e_i ) = { m
_{e_j} ^5 \over 384 \pi ^3 } ( {2\pi g_A^2 \over m_s^2 K_A })^2 [ 2
\vert \tilde g ^{LL}_{ji, ii} \vert ^2 + 2 \vert \tilde g ^{RR}_{ji,
ii} \vert ^2 + \vert \tilde g ^{LR}_{ji, ii} \vert ^2 + \vert \tilde g
^{RL}_{ji, ii} \vert ^2 ] , \eea where we have included an extra
symmetry factor $\ud $ in order to account for the pair of identical
charged leptons in the final states.

The bounds on the string scale implied by the mesons mass shifts and
the charged leptons decay rates are expressed by the explicit formulas
\bea && m_s \geq \bigg [ - {2\pi g_A ^2 \over K_A} {2m_P F_P^2 \over
\D m_P } [ {1 \over 3} \Re ( \tilde g^{LL} _{ij,ij} + \tilde g^{RR}
_{ij,ij} ) - \a ' _{ij} \Re (\tilde g^{LR} _{ij,ij} ) ] \bigg ]
^{1\over 2} , \cr && m_s \geq ( {2 \pi g_A ^2 \over K_A} ) ^ \ud \bigg
[ {m_j ^5 \over 384 \pi ^3 \G ( e_j \to e_i + e_i +\bar e_i ) } ( 2
\vert \tilde g ^{LL} _{ji, ii} \vert ^2 +2 \vert \tilde g ^{RR} _{ji,
ii } \vert ^2 + 2 \vert \tilde g ^{LR} _{ji, ii } \vert ^2 + 2 \vert
\tilde g ^{RL} _{ji, ii} \vert ^2 ) \bigg ] ^{1/4} . \eea In view of
the partial information that we dispose on the matrices $ V^f _H $ and
the complicated summations over the flavor basis amplitudes, we choose
to perform an approximate calculation motivated by the specific flavor
structure of contact interactions for the model at hand.  We only
retain the coefficients $ g_{ij,kl}^{HH'} $ with $ i=j, \ k=l $,
denoted by $ g ^{HH'}_{ik} \equiv g^{HH'}_{ii,kk}$, and neglect the
distinction between intersection points, by assuming the diagonal
terms $ g _{ii} $ to be independent of $i$ and the non-diagonal terms
to be symmetric, $ g_{ij} = g_{ji} $.  Applying now the unitarity
conditions on the quarks transformation matrices, $ V^q _H $, allows
us to express the mass basis coefficients for the neutral meson
observables in the form \bea && \tilde g^{HH'}_{ij,ij} \simeq (2
h_{ij} h ^{'}_{ij} + \tilde h _{ij} h '_{ij} + h_{ij} \tilde h ^{ '
}_{ij} ) (g^{HH'}_{ii} -g^{HH'}_{ik}) + 2 \tilde h _{ij} \tilde h ^{
'}_{ij} (g^{HH'}_{ii} -g^{HH'}_{jk}) + ( \tilde h _{ij} h '_{ij} +
h_{ij} \tilde h ^{ '}_{ij} ) (g^{HH'}_{ij} -g^{HH'}_{jk}) , \cr &&
[h_{ij} = V^ {q} _{ H ij} V ^{q \star }_{Hii} ,\ \tilde h_{ij} = V ^{
q\star }_{H ij} V^q_{Hjj} , \ i \ne j \ne k ] \eea where $ h '_{ij} ,
\ \tilde h '_{ij} $ are given by same formulas as $ h _{ij} , \ \tilde
h _{ij} $ with $ H\to H'$.  Assuming further that the non-diagonal
elements are independent of the specific pair, $i, j$, so that $g_{12}
= g_{13} = g_{23} $, leads to the factorized form \bea && \tilde
g^{HH'}_{ij,ij} \approx \calv ^P_{ij} (g^{HH'}_{ii} - g^{HH'}_{ij}) ,
\cr && [\calv ^P_{ij} = 2 h_{ij} h ^{'}_{ij} + 2 \tilde h _{ij} \tilde
h ^{ ' }_{ij} + \tilde h _{ij} h '_{ij} + h_{ij} \tilde h ^{'}_{ij} ]
. \label{eqmsshft} \eea

We have considered the alternative approximation defined by assuming
that the diagonal coefficients $ g_{ii} = g_d $ are independent of $i$
and the non-diagonal ones satisfy $ g_{ik} = g_{ki} \simeq g_{nd} $,
but without imposing the unitarity conditions.  The resulting form of
the $\D F=2$ coefficients reads \bea && \tilde g^{HH'}_{ij,ij} \simeq
V _{ij} ^{HH'} g_d + ( W _{ij} ^{HH'} + W _{ji} ^{H'H \star } + X
_{ij} ^{HH'} ) g_{nd} , \cr && [V _{ij} ^{HH'} = \sum _k V ^q_{H, ik}
V^{^q \star } _{H, jk} V^q_{ H', ik} V^{q \star } _{H', jk} , \ W
_{ij} ^{HH'} = \sum _{ k \ne l} V^q_{H, ik} V^{q \star } _{H, jl}
V^q_{H', ik} V^{q \star } _{H', jk} , \ X _{ij} ^{HH'} = \sum _{ k \ne
l} V^q_{H, ik} V^{q \star } _{H', jk} V^q _{H', il} V^{q \star } _{H,
jk} ] .
\label{eqmft2} \eea

For the coefficients $ \tilde g^{HH'}_{ij,ii} $ entering the charged
leptons decay widths, we use the same assumptions as in the
calculation done just above to obtain the simplified formula for the
mass basis coefficients \bea && \tilde g _{ji,ii} = \calv _{ji} g
_{ii} + \calv '_{ji} g _{ji} , \cr && [ \calv _{ji} = \sum _k V^{l}
_{Hjk} V^{l\star } _{H ik} V^{l } _{ H'ik} V^{ l\star } _{ H' ik} ,\
\calv '_{ji} = \sum _{k\ne l} V^{l} _{ H jk} V^{ l \star } _{ Hik}
V^{l } _{ H'il} V^{l \star } _{ H'il} ] .
\label{eqchglep}   \eea

\subsection{Results and discussion}
\label{subsect42}

We have numerically calculated the mesons mass shifts and the
three-body leptonic decay rates for Cremades et al.,~\cite{CIM,CIM2}
model with the reference set of parameters described previously and
the longitudinal distances set at, $ \e _A = (0,0,0) ,\ \e _B =
(0,1/3,0) $. Somewhat arbitrarily, we choose to set the various flavor
mixing matrices equal to the CKM matrix, $ V^{q,l} _H = V_{CKM},\
[H=L,R]$.  The following input data, expressed in GeV units, are
needed to calculate the mass shifts: \bea && m_u = 3 \times \ 10 ^{-3}
, \ m_d = 7 \times \ 10 ^{-3} , \ m_s = 0.095 , \ \ m_c = 1.25 , \ m_b
= 4.20 ; \ \cr && F_K = 0.1598 , \ m_K = 0.497 , \ \D m _K= 3. 483 \
\times \ 10 ^{ - 15} , \ \calv ^K = 0.0979 ; \cr && \ F_B = 0.176 , \
m_B = 5.2794 , \ \D m _B= 3.337 \ \times \ 10 ^{ - 13}, \ \calv ^
{B_d} = 0.000125; \cr && \ F_{B_s} = 0.176 , \ m_{B_s} = 5.367 , \ \D
m _ {B_s}= 1.145 \ \times \ 10 ^{ - 11} , \ \calv ^ {B_s} = 0.00347 ;
\cr && \ F_{D} = 0.2226 , \ m_{D} = 1.8645 , \ \D m _ {D}= 4.607 \
\times \ 10 ^{ - 14} , \ \calv ^ {D}=0.0979 .
\label{eqmesons}\eea The following input data, expressed in GeV units,
are needed to calculate the charged leptons three-body decays: \bea &&
m_e = 0.511 \times \ 10 ^{-3} , \ m_\mu = 0.1056 , \ \G (\mu \to e + e
+ \bar e ) < 3.29 \ \times \ 10 ^{-31} , \ \Re \calv _{21}= -0.198633
,\ \Re \calv ' _{21} = 0.198722 ;\cr && m_\tau = 1.777 , \ \G (\tau
\to e + e + \bar e ) < 4.529 \ \times 10^{-19} ,\ \Re (\calv _{31})=
0.00641 ,\ \Re (\calv ' _{31}) = -0.00660 . \label{eqleptons}\eea

The results obtained with the flavor mixing described by the
approximate formulas in Eqs.~(\ref{eqmsshft}) and (\ref{eqchglep}) are
plotted in Fig.~\ref{fig7}.  The use of Eq.~(\ref{eqmft2}) gives
similar results.  We see that the bounds on $ m_s $, at fixed $ m_s
r$, increase with increasing $ m_sr$ according to the approximate
power law, $ (m_s r) ^ {5/4} .$ Wide disparities appear between
different cases mainly because of the flavor mixing factor. The most
constraining observable, corresponding to the $ K-\bar K$ mass shift,
yields the bound, $ m_s > O(10^3 ) $ TeV.  Relaxing our assumption
that the Standard Model contributions are negligible can only
strengthen the bounds on $m_s$.  However, one may expect significantly
weaker bounds if the flavor and mass bases were not too strongly
misaligned so that the flavor change is dominated by the direct
contributions.  For instance, using the order of magnitude predictions
for the off-diagonal coefficient $ g_{ij,kl} ^{HH'}$ in panel $(c)$ of
Fig.~\ref{fig1}, with $ V_H^f =1$, would reduce the bound on $m_s $
from the $ K-\bar K$ mass shift by a factor of order $10^{-1} \ - \
10^{-2}.$ A careful treatment of the flavor structure of the model
would be needed in order to make a more definite statement.

For comparison, we note that the bound from the $ K \ - \ \bar K$ mass
shift obtained in~\cite{abel03,abeleb03} by using the approximate
representation of Eq.~(\ref{eqloexp}) for the string momentum modes
reads, $ \ m_s \geq 100 $ TeV.  A similar gap exists for the other
flavor changing observables.  However, these results were obtained by
setting, $ m_s r \simeq 20$, which lies well above the allowed
interval.

We comment briefly on the CP violating observable, $\e _K$, which is
set by the experimental data for the $ K_L ^0 \to \pi + \pi $ decays
to the value, $ \vert \e _K \vert _{exp} \simeq \vert \eta _{00} \vert
= (2.285 \pm 0.019 ) \ 10 ^{-3} $.  Since the coefficients $
g^{HH'}_{ij,kl}$ are real, the prediction for $ \e _K$ depends in a
crucial way on the flavor mixing matrices.  In our treatment of the
indirect flavor mixing leading to Eq.~(\ref{eqmsshft}), the
predictions for $\e _K$ and the mass shift scale as, $ \vert \e _K /
\D m_K\vert = \vert \Im ( \calv ^K ) / \Re ( \calv ^K ) \vert $.
Since the CP violation effects enter the CKM matrix through the second
and third fermion generations, $\calv ^K_{12} $ is real and hence $\e
_K =0$.  The alternative prescription for the flavor mixing described
by Eq.~(\ref{eqmft2}), with the matrices $ V_H^f $ still identified
with the CKM matrix, yields an uninteresting small bound on $ m_s $.

It is also instructive to compare with the split fermion models.  The
bound from the $ K-\bar K $ mass shift~\cite{lill03}, $ M_c > \b
[\calv ^K F (\rho \a ) ]^\ud \simeq (100\ -\ 600 ) $ TeV with $\b =
1125 $ TeV and $ F (\rho \a )\simeq (2. \ - 8. )$, is of same
magnitude as ours, while sampling over the parameters which control
the indirect flavor mixing effects give the ability to suppress the
bound by factors of $ 10 \ - \ 100$.  The description of exchange
contributions in these models differs from that in intersecting brane
models where the flavor hierarchies originate in the instanton
contributions, rather than the wave function overlaps, and the
parameter space is more restricted.  In addition to the extra
dimension size parameter, $M_c =1/r $, the split fermion
models~\cite{lill03} introduce the scaled localization width and
fermion separation parameters, $ \rho = \s / r $ and $ \a = \D y / \s
$, which qualitatively identify with the string theory parameters, $
\rho \sim M_c /m_s $ and $\a \sim \e m_s / M_c = \e m_s r$, assuming $
\s \sim 1 /m_s$.  We conclude from this indirect comparison that the
wide hierarchies, $ \a \in [0, 15]$ and $\rho \in [10^{-1}, 10^{-4} ]
$, which are needed in split fermion models to weaken the bounds on $
M_c$, are not favored by the analogous intersecting brane models.

Finally, we present the result of an indicative study of the
tau-lepton hadronic and semi-hadronic decays, $ \tau \to \pi + \mu $
and $\tau \to \pi + \pi + \mu $, based on the analysis~\cite{black03}
of the effective interaction for the associated subprocesses, $ \D L
_{EFF} = {4 \pi \over \L ^ {\tau q 2 } } (\bar \mu \g ^\mu \tau )
(\bar q \g _\mu q ) + H. c.  $, which yields the bound, $\L ^{\tau q }
> 12 $ TeV. Using the predictions in panel $(c)$ of Fig.~\ref{fig1},
we deduce the bound $m_s = \L ^ {\tau q }g_a (g _{lq} ^{LL} )^\ud >
0.5 $ TeV at $ m_s r=1$.  At larger $m_s r$, neglecting the flavor
mixing effects leads to useless small bounds due to the strong
suppression of the flavor non-diagonal string amplitude.

\section{Summary and Conclusions}

We have discussed in this work collider and flavor physics tests of
the four fermion tree level string amplitudes in intersecting brane
models.  Although the study was specialized to the isolated
orientifold premodel of Cremades et al.,~\cite{CIM,CIM2} realizing the
Standard Model, this is a good representative of the families of
string models selected in current explorations of the landscape of
open string vacua.  Based on a qualitative examination of the
predictions for the gauge coupling constants, we also verified that it
is compatible with a TeV string mass scale.

The string theory predictions depend on two free mass parameters, $
m_s $ and $ M_c =1/r $, along with the geometric shape parameters of
the internal $ T^6$ torus and known inputs for the electroweak gauge
bosons masses and gauge coupling constants.  The necessary condition
for weakly coupled open strings imposes the restricted variation
interval, $ m_s r = m_s /M_c \in [1,4]$.  We have studied the four
fermion contact interactions from exchange of string Regge and winding
modes in various configurations of the quarks and leptons, paying
special attention to the gauge group structure and the contributions
from world sheet instantons.  The general features of predictions for
the contact interactions may be briefly summarized as follows.  The
size of coefficients present regularities which reflect in part the
symmetric configuration of the brane setup.  The sensitivity of
predictions to the tori shape parameters leads to a moderate
sensitivity of the flavor diagonal coefficients $\D F=0$ on geometric
parameters which spreads predictions by a factor $ 2\ - \ 3$.  The
widest disparities occur between the mixed and unmixed chirality
amplitudes.  Two characteristic features reside in the strong growth
of the flavor diagonal coefficients, $ (m_s r ) ^{5/2} $, and the
strong suppression of the flavor non-diagonal $\D F=1, \ 2 $ relative
to $\D F=0$ by factors of order $ 10^{-1} $ and $ 10^{-4}$, due to the
classical partition function factor when the distances between
intersection points relative to the wrapped cycles radius are of $ \e
= O(1/3)$.

The Bhabha scattering differential cross section is an important high
precision observable for which the theoretical and experimental
uncertainties are expected to reach $ O( 10^{-3}) $ in the future.  We
have considered a qualitative comparison with the LEP data which leads
to bounds on the string mass scale of TeV order.  These are expectedly
stronger than the bounds obtained in the single brane
model~\cite{cpp00} where the local operators have dimension $8$.  It
should be useful to pursue a systematic study for the set of $2-2 $
body processes including the Drell-Yan lepton pair production and the
parton subprocesses with initial states, for $e+q ,\ q +\bar q '$ and
$ q+q'$.

We have also considered the direct contributions to the four fermion
contact interactions from string Regge and winding modes to a subset
of flavor changing neutral current observables using an approximate
description of the indirect flavor mixing effects where the direct and
indirect flavor changing effects factorize.  The $ K -\bar K $ mass
splitting yield the strongest constraint, $ m_s > 10^3 $ TeV.  This
bound, as well as other ones deduced from flavor changing observables,
are an order of magnitude stronger than those obtained from the
contributions to contact interactions due to the string momentum
modes~\cite{abel03,abeleb03}.  It is fair to say, however, that this
conclusion is at best qualitative since the two calculations rely on
different inputs and approximations.  To obtain more realistic
estimates, the highest priority should be set on obtaining realistic
inputs for the flavor mixing matrices which match the predictions for
the fermions mass matrices to observations while improving on the
restrictive rank 1 property of the fermions mass matrices in the model
at hand.

\appendix

\section{Brief review of open string sectors in toroidal orientifolds} 
\label{sappen0}

     We consider type $II$ string theory compactified on factorisable
toroidal orientifolds, $T^6 / \O \calr ,\ [T^6=\prod _{I =1}^3 T^2_I
]$ where the involution symmetry, $\calr =\prod _{I =1}^3 \calr _I$,
acts on the orthogonal basis of complex coordinates, $ X^I= (X^I_1 +i
X^I_2)/\sqrt 2$ in the $ T^2_I,\ [I=1,2,3]$ complex planes as
reflections across the real axes, $\calr _I \cdot X ^I = \bar X^I $.
We restrict to the subset of factorisable three-cycles in the integer
homology vector space, $\Pi _\mu \in H^3 (T^6, Z) $, represented in
terms of the homology of one-cycles with lattice dual bases, $[a^I], \
[b^I] \in H^1 (T^2_I, Z) $, by the three pairs of integer quantized
winding numbers, $ (n^I_\mu , \ m^I_\mu ) $. These three-cycles are
represented in the orthogonal coordinate system of the three complex
planes of $ T^6$ by \bea && [\Pi _\mu ]= (n^I_\mu , \ \tilde m^I_\mu )
, \ [ \tilde m^I_\mu = m^I_\mu - n^I_\mu U_1^I , \ U ^I \equiv U^I_1 +
i U^I_2 = - {e_1 ^I \over e_2 ^I} = - \hat b_I + {i \over \chi ^I },\
\chi _I = {r_2^I \over r_1 ^I} ]
\label{eqtori} \eea 
where $ U^I$ denote the $ T^2_I$ tori complex structure moduli whose
real parts are subject to the restriction, $\hat b_I = 0,\ \ud $, for
orthogonal and tilted tori, respectively. We have denoted by $ r_1
^I,\ r_2 ^I$ the radius parameters of the one-cycles of $ T^2_I$
projected on the pair of orthogonal axes.  For non-orthogonal 2-d
tori, $T^2_I$, we choose to work with the case of upwards tilted tori,
where $ r_2 ^I $ refers to the one-cycle along the imaginary
(vertical) axis of the complex plane and $ r_1 ^I $ to the projection
of the dual one-cycle radius along the real (horizontal) axis.  The
orientifold $O6$-planes are the loci of points fixed under $\calr $
which extend along the three uncompactified dimensions of Minkowski
space-time, $M_4$, and wrap the three one-cycles, $ (n^I_\mu , m^I_\mu
) = (1,0)$.  Both the $O6$-planes and $D6$-branes are sources for the
closed string RR modes seven-form, $ C_7$, with RR charges determined
by the winding numbers of the wrapped three-cycles. The divergent
tadpoles of RR modes due to the $O6$-planes in the one-loop closed
string (Klein bottle surface) amplitude are assumed to cancel against
the tadpoles in the open string (cylinder and M\"obius strip surface)
amplitudes contributed by introducing $K$ parallel stacks of $N_\mu $
branes $D6_\mu ,\ [\mu = 1, 2, \cdots , K = a, b, \cdots ]$.  To the
$D6_\mu $-brane stack wrapped around $ [\Pi ^I_{\mu }] = ( n_{\mu } ^I
, \tilde m_{\mu }^I)$, is associated the orientifold mirror image
$D6_{\mu '}$-brane stack, wrapped around the image cycle $ [\Pi
^I_{\mu '}] = ( n_{\mu '} ^I , \tilde m_{\mu '}^I) = ( n_\mu ^I ,
-\tilde m_\mu ^I) $. For toroidal orientifolds, the RR tadpole
cancellation conditions are of form \bea && \sum _{\mu =1} ^K N_\mu n
^I_\mu n ^J_\mu n ^K_\mu + \ud Q_{Op} =0 , \ \sum _{\mu =1} ^K N_\mu n
^I_\mu \tilde m ^J_\mu \tilde m ^K_\mu = 0 , \ [Q_{Op} = \mp 2 ^{p-4}
f_p = \mp 32 ] \eea where the summations run over the orientifold
equivalence classes, counting mirror pairs as single units; $ I, J, K
\in [1,2,3] $ run over the distinct permutations of the complex planes
indices; $f_p = 2 ^{9-p} $ denotes the $Op$-planes multiplicity; and
the upper and lower signs of the orientifold charge $ Q_{Op}$ refer to
the $SO$ and $Sp$ (orthogonal and symplectic group) orientifold
projections.

The $D6_a $-brane location in the $T^2_I$ complex planes is described
by an oriented vector tilted relative to the $O6$-plane (along the
real axes) by the angles, $\pi \t _{a } ^I = \arctan (\tilde m ^I _ a
/n^I _ a U_2 ^I ) $.  The $D6$-branes serve as boundaries for the end
points of open strings which carry their perturbative excitations.
The open string sectors, $(a,b)$ and $(a,b')$, associated to the two
pairs of branes, $D6_a/D6_b $ and $D6_a/D6_{b'}$, are assigned the
interbrane angles, $ \t _{a b} = \t _b - \t _a $ and $ \t _{a b'} = \t
_{b'} - \t _a = -\t _{b} - \t _a $.  We use notational conventions
where the brane-orientifold and interbrane angles vary inside the
intervals, $\t _{a ,b } ^I \in [-1,+1] $ and $\t _{a b} ^I \in
[0,+1],\ [I=1,2,3]$ with positive sign angles associated to
counterclockwise rotations. Transforming back to values of the angles
inside these ranges requires geometric information on the signs of
winding numbers.  The brane pairs $a,\ b $ intersect at fixed numbers
of points determined by the topological invariants \bea && I_{a b} =
\prod _I (n_a^I \tilde m_b^I -\tilde m_a^I n_b^I ) ,\ I_{a' b} = \prod
_I (n_{a'} ^I \tilde m_{b}^I -\tilde m_{a'} ^I n_{b} ^I ) = \prod _I
(n_a ^I \tilde m_b^I + \tilde m_a ^I n_b^I ) . \eea

The low-energy dynamics on a single isolated stack of $ N \ D6$-branes
in the 4-d space-time $M_4$ is approximately that of a gauge field
theory with gauge group $U(N)$, supersymmetry $\caln =4$, and a
certain content of massless modes associated with the branes moduli.
The open string sectors for the $D6_a /D6_b $-brane pair supporting
the gauge symmetry $ U(N_a) \times U(N_b) $ include: (1) The diagonal
modes, $(\mu ,\mu ), \ [\mu =a, \ b]$ which carry the adjoint
representations; (2) The orientifold twisted modes, $(\mu ,\mu ')$,
which carry the symmetric and antisymmetric representations $\bfA ,\
\bfS $ of $ U(N_\mu ) $ with the multiplicities, $ \ud (I _{\mu \mu '}
\pm I _{\mu O6} ) , \ [I _{\mu O6} \equiv [\pi _\mu ] \cdot [\pi _
{O6}] = \prod _I(- m_\mu ) ] $; and (3) The non-diagonal `twisted'
modes, $(a, b ) \sim (b,a) ^\dagger $ and $(a', b ) \sim (b,a')
^\dagger $, which carry the bifundamental representations, $ I_{ab} (
N_a , \bar N_b) \oplus I_{a'b}(\bar N_a ,\bar N_b) $.  The equivalence
relations between open string sector sectors, $ (a,b) \sim (b',a')\sim
(b,a) ^\dagger ,\ (a,b') \sim (b,a') \sim (b',a) ^\dagger $, where the
dagger stands for the complex conjugation of the states space-time and
internal group quantum numbers, lead to interpret the intersection
numbers $ I_{a b}$ or $ I_{a 'b}$ of negative signs as multiplicities
for the modes with conjugate chirality and group representation, $
\vert I_{a b} \vert (\bar N_a , N_b)$ or $ \vert I_{a' b} \vert (N_a ,
N_b)$.

The key condition to preserve $\caln =1$ supersymmetry on the branes
4-d intersection is that the wrapped three-cycles be of special
Lagrangian type. For Calabi-Yau manifolds, these are the cycles whose
real number valued volume integrals are calibrated by the holomorphic
three-form, $ e^{i\varphi } \O _3 $, characterized by a fixed choice
of the angle parameter, $\varphi $.  In close analogy with the
conditions selecting in closed string theories the holonomy of
subgroups of the internal space manifold symmetry group $SO(6)$, one
preserves $\caln =1 $ or $ \caln =2$ supersymmetry to the extent that
the rotation matrices relating the branes to the orientifold planes
belong to the group $ SU(3)$ or $ SU(2)$~\cite{berk96}.  The
requirement that the wrapped cycles $\Pi _a $ and $ \Pi _b$ are
calibrated by the same three-form so that the $D6_a /D6_b $-brane pair
preserves $\caln =1$ supersymmetry, amounts to the conditions, $\sum
_I \t _{a, \ b} ^I =0\ \text{mod} \ 2 $.  In terms of the spinor
weights $ r ^a _{(\a )} $ of $ SO(8)$ for the $16$ supercharges
conserved in the bulk, $\caln =1, \ 2$ supersymmetry arises when a
single or a pair of spinor weights $r _{(\a )} $ solves the equations,
$ \sum _{a =1} ^4 r _{(\a )} ^a \t ^a _\mu = 0,\ [a=1, \cdots , 4;\
\mu =1, \cdots , K]$ where the intersection angle in $M_4$ is set here
to zero, $ \t ^4 =0$. In the basis of independent spinor charges for $
SO(8)$ defined by \bea && r_{(1)} = (-++-), \ r_ {(2 )} = (+-+-), \
r_{(3 )} = (++--), \ r_{(4 )} = (----),\eea with the notational
convention, $ r _{(1 )} ^a = (-++-)= (-\ud , \ud , \ud , -\ud ) , \
[a=1,2,3,4]$ the three special angle configurations, $ (0,\t _A, \pm
\t _A),\ (\t _B, 0, \pm \t _B),\ (\t _C, \pm \t _C,0),$ preserve the
$\caln =2$ supersymmetries associated to the pairs of spinor weights,
$ {r_{(2 )} , r_{(3 )} \choose r_{( 1)}, r_{(4 )}} ,\ {r_{(1 )} , r_{(
3)} \choose r_{(2 )}, r_{(4 )}} ,\ {r_{(1)} , r_{(2)} \choose r_{(3)},
r_{(4)} } $.

We discuss next the dependence of intersection points on geometrical
data for the angles and transverse separations of $ D6_A/D6_B$-brane
pairs that do not necessarily pass through a common point of $T^6$.
Suppressing the index $I$ of the 2-d tori $T^2 _I$, for convenience,
we parameterize the $D6_A$-brane wrapped around the one-cycle of $
T^2$ with winding numbers $ (n _A , m _A )$ by the equation \bea &&
X_A (\xi _A) = ( L_A \xi _A + q_A + p_A \tau +d _A t_A ) e_1, \eea
where \bea && L_A = n_A + m_A \tau =\tilde n_A + i m_A \tau _2 , \ t
_A = -m_A \tau _2 + i \tilde n_A ,\ \tilde n_A = n_A + m_A \tau _1,
\cr && [\tau = \tau _1 +i\tau _2 = {e_2 \over e_1 },\ e_1 = 2\pi r_1,
\ e_2= 2\pi r_2 e ^{i\a } ] \eea with $\tau $ denoting the $T^2$ torus
complex structure modulus, $d _{A} $ the transverse displacement from
the origin, $ q_{A} ,\ p_{A} \in Z $ the lattice displacements along
the basis of dual cycles, and $\xi _A ,\ d _A \in R $ parameterize
points along longitudinal and transverse directions.  We have
formulated the problem here in the case of sideways tilted torus,
which is related to the case of upwards tilted torus considered in
Eq.~(\ref{eqtori}) by the transformation, $\tau = -1/ U $.  (The
formulas for the upwards and sideways tilted tori are also related by
the substitutions, $ n\to m ,\ m\to -n ,\ e_1\to e_2,\ e_2\to -e_1.$)
We now introduce a similar equation for $D6_B:\ X_B (\xi _B) / e_1=
L_B \xi _B + q_B + p_B \tau +d _B t_B $, and require the condition, $
X_A (\xi _A) =X_B (\xi _B) $. The resulting pair of linear equations
for the real variables $\xi _A $ and $\xi _B$, \bea && \pmatrix
{\tilde n_A & -\tilde n_B \cr m_A & - m_B } \pmatrix {\xi_A \cr
\xi_B}= - \pmatrix { Q_{AB} \cr P_{AB} } , \eea where \bea && Q_{AB} =
Q_{A} - Q_B,\ P_{AB} = P_{A} -P_B , \ Q_{\mu} = q_{\mu } - d_{\mu }
m_{\mu } \tau _2 , \ P_{\mu } = p_{\mu } + d_{\mu } \tilde n_{\mu } /
\tau _2 ,\ [\mu = A, B] \eea is solved in matrix notation by \bea &&
\pmatrix {\xi_A \cr \xi_B}= - {1\over I_{AB} } \pmatrix { m_B &
-\tilde n_B \cr m_A & - \tilde n_A } \pmatrix { Q_{AB} \cr P_{AB} } \
\Longrightarrow \ X _A = {k _{AB} L_A \over I_{AB} } , \ X _B = { k
'_{BA} L_B \over I_{AB} } , \cr && [I_{AB}= n _A m _B-m _A n_B ,\ k
_{AB} = (\tilde n_B P_{AB} - m_B Q_{AB}) \ \text{mod} ( I_{AB}) ,\
k'_{BA} = (\tilde n_A P_{AB} - m_A Q_{AB} ) \ \text{mod} ( I_{AB})]
. \label{eqshifts} \eea
The $ I_{AB}$ solutions for $ \xi _A ,\ \xi _B$ are in one-to-one
correspondence with the pairs of integers $ k _{AB},\ k'_{BA}$.  The
shift vectors, $ w_{BA} \equiv X_B -X_A = (k'_{BA} L_B - k_{AB} L_A )
/ I_{AB} $, which link the positions of a given intersection point
along the pair of intersecting branes $ ( B, A)$ in the complex plane
of $ T^2 $, belong to the lattice coset, $\L _{BA} / \L $, where $ \L
_{BA} $ denotes the grand torus lattice generated by the cycles $
L_B,\ L_A$, and $ \L $ the $ T^2 $ torus lattice~\cite{higaki05}.

We next discuss the vector space of open string states, $ (a,b) ,\
[a=b,\ a\ne b] $. The state vectors, $ \vert (a,b) k, N, r _{(\a )} ,
(A, ij ) > \l ^{(ab)}_{A, ij} $, are described by the four momentum
$k$; the coordinates oscillator number $ N $; the weight vector $ r
_{(\a )} $ of the $ SO(8) $ group Cartan torus lattice; and the gauge
group multiplet component $A$. The CP gauge factors, $\l ^{(ab)} _{A,
ij} , \ [i =1, \cdots , N_a; \ j =1, \cdots , N_b] $ are matrices
whose array and column indices $ i = (1, \cdots , N_a ) , \ j = (1,
\cdots , N_b )$ label the coincident branes inside the stacks $ a, \
b$ of size $ N_a, \ N_b$.  In orientifolds, the combined systems of
mirror brane pairs, $ (a + a', \ b + b' )$, are described by single
matrices, $\l ^{(ab)} _{A, ij}, \ [i , \ j \in ( a + a ' ,\ b + b ') ]
$ of size $ (2N_a + 2 N_b) \times (2N_a + 2 N_b) $. These are
conveniently represented by $ 2\times 2 $ block matrices with
sub-blocks labeled by $ a , a' , b, b' $.  The diagonal open string
states $(a,a)$ are ascribed $ 2N_a \times 2 N_a $ gauge matrices, $\l
^{(aa)} _{A, ij}$.  We drop in the following the suffix labels $ (a
b),\ (aa) $ and $ A , \ i, j $ on the matrices, except when needed.
We use conventions in which the matrices satisfy the normalization and
closure sum conditions, $ Trace(\l _A \l_B ^ \dagger ) = \d _{AB}
,\quad \sum _A Trace(O_1 \l _A ) Trace(O_2 \l _A ) = Trace(O_1 O_2). $

The orientifold symmetry $\O \calr $ acts on the gauge quantum numbers
of brane stacks through the twist matrix given by the direct product
of $ 2N_\mu \times 2 N_ \mu $ unitary matrices, $\g _{\O \calr , \mu }
,\ [ \mu = a, b ]$. The orientifold projection on physical states is
then defined by, $ \l ^{(ab)} _A = \eta _A \g _{\O \calr ,a} \l ^{(ab)
T } _A \g _{\O \calr ,b} ^{-1} , $ where $ \eta _A $ are state
dependent complex phase factors.  These conditions must be imposed
only for brane stacks $ \mu $ which wrap cycles coinciding with their
orientifold mirror images, $ \mu = \mu '$, hence fixed under the
orientifold twist, $ \Pi _{\mu } = \Pi _{\mu ' } = \Pi _{O6}$.  In the
case of a brane stack $ \mu $ at generic angles, $ \mu \ne \mu '$, no
conditions need be imposed beyond requiring that the representations
for $ \mu $ and $ \mu '$ be conjugate.  Nevertheless, it is convenient
to treat in a unified way the cases with $\mu = \mu ' $ and $\mu \ne
\mu ' $, by taking the orientifold projection matrix in the latter
subspaces to be trivial, $\g _{\O \calr ,\mu } =1$.  The $ SO$ and $
Sp$ orientifold projections, corresponding to $ Q _{O6} =\mp 2 ^{5} $,
are characterized by the property of the twist embedding matrix, $ \g
_{\O \calr ,\mu } ^T = \pm \g _{\O \calr , \mu } $.

We now specialize to the $Sp$ type projection which is the appropriate
one for the model of interest to us to be discussed in
Subsection~\ref{subsect23}.  Solving the condition for the
bifundamental representation, $\l ^{(\mu \nu )} = - \g _{\O \calr ,\mu
} \l ^{(\mu \nu ) T} \g _{\O \calr , \nu } ^{-1} $, with antisymmetric
twist matrices, $ \g _{\O \calr , a } , \ [ a= \mu ,\nu ]$ of
dimension $ (2 N_\mu + 2 N_\nu ) \times (2 N_\mu + 2 N_\nu ) $, and
the similar condition for the adjoint representation matrix, $ \l
^{(G_\mu )} $, of dimension $ 2 N_\mu \times 2 N_\mu $, one obtains
\bea && \l ^{(G_\mu )} = \pmatrix{ m & s_1\cr s_2 & - m^T} , \ \l ^{
(\mu \nu ) } = \pmatrix{ 0 & B \cr B ' & 0 } ,\cr && [B= \pmatrix{\a &
\b \cr \g & \d } , \ B'= \pmatrix{-\d ^T & \b ^T \cr \g ^T & - \a^T}
,\ \g _{\O \calr ,\mu } = \pmatrix{0 & 1 _{N_\mu } \cr - 1 _{N_\mu } &
0} ]
\label{eqcp} \eea 
where the sub-blocks $ m ,\a ,\ \b , \ \g , \ \d $ and $ s _1 , \ s_2
$ designate arbitrary generic and symmetric matrices with $ m, \ s _1
, \ s_2 $ and $ \a ,\ \b , \ \g , \ \d $ having dimensions $ N_\mu
\times N_\mu $ and $ N_\mu \times N_\nu $, respectively.  The
conjugate group representations are assigned Hermitian conjugate
matrices.  The adjoint representation matrix $ \l ^{ (G_\mu )} $ in
Eq.~(\ref{eqcp}) involves $ 2 N_\mu ( 2N_\mu + 1) /2 $ independent
parameters, as needed to match the dimension of the gauge group $ USp(
2N_\mu )$.  The special case for the matrix $ \l ^{ (G _\mu ) } $ with
$ s_1=s_2 =0$ corresponds to the adjoint representation of $U(N_\mu
)$, involving the expected number of $ N_\mu ^2 $ independent
parameters.  For the bifundamental representation matrix $\l ^{ (\mu
\nu )} $, the entries in the sub-blocks $ B$ and $ B' $ are related by
the requirement that the substitution $\mu \leftrightarrow \mu '$
corresponds to charge conjugation.  The infinitesimal transformations
of the gauge matrices $ \l ^{ (R)} $ in the representation $R$ of
sector $(\mu , \nu ) $, obtained from the commutator with the adjoint
representation matrix, \bea && \d _{\mu \nu } \l ^{(R)} \equiv ( \d
_\mu + \d _\nu ) \l ^{ (R)} \equiv [ \l ^{( G_\mu ) } \oplus \l ^{
(G_\nu )} , \l ^{(R)} ] , \qquad [\l ^ {(G_\mu )} = ( \e _ \mu , - \e
_ \mu ) \otimes I_\nu + I_\mu \otimes ( \e _ \nu , - \e _ \nu ) ] \eea
act on the sub-block matrix entries of the matrices $B,\ B'$ for the
bifundamental representations as \bea && \d _{\mu \nu } \a = \e _\mu
\a -\a \e _\nu ,\ \d _{\mu \nu } \b = \e _\mu \b + \b \e _\nu ,\ \d
_{\mu \nu } \g = -\e _\mu \g -\g \e _\nu ,\ \d _{\mu \nu } \d = - \e
_\mu \d + \d \e _\nu . \eea The four inequivalent bifundamental
representations are thus in one-to-one correspondence with the
sub-blocks, $\a , \ \b ,\ \g , \ \d $, characterized by the charge
assignments, $ (Q_\mu , Q _ \nu ) : \a \sim (1,-1),\ \b \sim (1,1),\
\g \sim (-1,-1),\ \d \sim (-1,1).$


\section{Two-body processes at high energy colliders}
\label{sectappen1}

\subsection{Tree level string amplitudes for processes with fermion 
and gauge boson pairs}
\label{subsecapp1}

We discuss in this appendix the tree level four point open string
amplitudes in models related to the intersecting brane models.  For
comparison with the results in Subsec.~\ref{subsect22}, we first
consider the branes within branes models. We start, for completeness,
with the case of four fermion modes belonging to the diagonal sector
$(p,p)$ of a $ Dp$-brane which corresponds to the single $ Dp$-brane
model~\cite{cpp00}. The result can be derived by dimensional reduction
of the familiar formula for the $ D9$-branes of type I theory \bea &&
\cala ' _{(p,p)^4} = G_{ Dp} \calT _{1234} \int _0^1 dx x ^{-s -1}
(1-x )^{-t -1} [(1-x) \cals _{1234} - x \cals _{1423} ]+ \text{perms}
\cr && = -G_{ Dp} \calT _{1234} { \cals (s,t) \over st } (t \cals
_{1234} - s \cals _{1432} ) + \text{perms} , \eea where we use same
notations as in Eqs.~(\ref{eqstramp1}) and~(\ref{eqamp4}).  The above
formula is formally related to that in Eq.~(\ref{eqstramp1}) by the
substitution, $ C \cals _{1234} \calv _{1234} \to - G_{ Dp} (t \cals
_{1234} -s \cals _{1432}) {\cals (s,t) \over s t }.$ Matching to the
pole term from massless gauge boson exchange determines the
normalization factor as, $ G_{ Dp} \calT _{1234} = 2 g_{Dp} ^2 $,
where the $Dp$-brane gauge coupling constant $g_{Dp} $ enters the
gauge current vertex as, $ \sqrt 2 g_{Dp} T^a \g _\l $.

We now focus on the massless fermion modes of the $ Dp/D(p+4)$-brane
models localized at brane intersections in the non-diagonal sectors,
$(p,p+4) + (p+4,p)$.  The calculations, discussed initially
in~\cite{antbengier01}, bear formal similarities with those for
intersecting branes.  We specialize to the case $ p=3$ of setups with
$ D3/D7_I$-brane pairs with $I=1,2,3.$ There are two possible channels
for the couplings of four non-diagonal sector modes $(3,7_I) $, which
correspond to the configurations $ \vert (3, 7)_I\vert ^4$ and $ \vert
(3, 7_I)\vert ^2\vert (3, 7_J)\vert ^2 $.  The string amplitude for
the first channel, $ [(3, 7)_I ]^4$, involving two identical pairs of
conjugate fermion modes is given by \bea && \cala ' _{ (3,7_I )^4} =
C' _{Dp} \int _0^1 dx x ^{-s-1} (1- x )^{-t-1} [ (1-x) \cals_{1234} -
x \cals_{1432} ] \bigg ( T _{1234} \prod _{A } ( { \vt [ { \e _A \atop
0} ] (\tau _A ) \over F ^ \ud (x) } ) + T_{4321} \prod _{A } ( { \vt [
{ \e _A \atop 0} ] (\tau _B ) \over F ^ \ud (1-x) } ) \bigg ) ,\cr &&
[ \tau _A (x) = i \vert L_A \vert ^2 { F(1-x) \over F(x) } , \ \tau
_B(x) = i \vert L _A\vert ^2 { F(x) \over F(1-x) } ,\ F(x) = F (\ud
,\ud ; 1; x) ] \eea where the label $ A= (J_m, \ K_m) ,\ [m=1,2] $ in
the products runs over the real dimensions of the 4-d sub-torus $ T^2
_J \times T^2 _K$ of the internal torus $ T^6$ wrapped by the $
D7_I$-brane, the Wilson line parameters along the corresponding
sub-torus are denoted by $ \e _A$, and the sub-torus volume parameter
is defined by, $ \vert L_A \vert ^2 = r_J^2 r_K^2 $.  The direct and
reverse orientation terms inside the large parentheses are related by
the change of integration variable, $ x \to (1-x)$.  Combining the
regions of the $x$-integral near $ x=0$ and $ x=1$ yields the
low-energy approximate representation of the string amplitude as
infinite series of $s$-channel or $t$-channel poles located at the
string compactification modes \bea && \cala ' _{ (3,7_I )^4 , 0} +
\cala ' _{ (3,7_I )^4 , 1} \simeq C' _{Dp} \sum _{p_A \in Z} \bigg (
T_{1234} [ \cals_{1234} { \prod _A \d ^{-(p_A + \e _A )^2 r_A ^2}
\over -s + \sum _{A} (p _A + \e _A )^2 r_A ^2 } - \cals_{4321} { \prod
_A \d ^{-p _A^2 / r^2 _A } e ^{2 \pi i p_A \e _A }/ r_A \over -t +
\sum _ {A} {p _A^2 / r_A ^2 } } ] \cr && + T _{4321} [ \cals_{1234} {
\prod _A \d ^{-p _A^2 / r^2 _A } e ^{2 \pi i p_A \e _A }/ r_A \over -s
+ \sum _ {A} {p _A^2 / r_A ^2 } } - \cals_{4321} { \prod _A \d ^{-(p_A
+ \e _A )^2 r_A ^2} \over -t + \sum _{A} (p _A + \e _A )^2 r_A ^2 } ]
\bigg ) ,\eea where the momentum modes in the open string sector
$(7_I,7_I)$ arise after use of the Poisson resummation formula and the
winding modes belong to the open string sector $(3,3)$.  The massless
pole terms determine the normalization constant in terms of the gauge
coupling constants by the formula, $C' _{Dp} T_{1234} = 2 \pi g_s =
{2g_{Dp} ^2 V_{p-3} m_s ^{p-3} \over (2 \pi ) ^{p-3} } $ with $
V_{p-3} \sim (2 \pi r ) ^{p-3} $ denoting the volume of the $
(p-3)$-cycle of the internal manifold wrapped by the $Dp$-brane.  For
the Abelian gauge group case, $ C' _{Dp} T_{1234} = C' _{Dp} T_{4321}
= 2 g_ {D3} ^2 = 2 g_ {D7} ^2 r_J ^2 r_K ^2 $.

For the channel $ \vert (3, 7_I) \vert^2 \vert(3, 7_J)\vert ^2 $,
involving two distinct pairs of conjugate fermion modes, $ (3, 7_I) $
and $ (3, 7_J) ,\ [I\ne J]$ the string amplitude is given by the
formula \bea && \cala ' _{ (3, 7_I) ^2 (3, 7_J) ^2 } = C' _{Dp} \int
_0^1 dx x ^{-s-1} (1- x )^{-t-1} \bigg ( T_{1234} [ (1-x) \cals_{1234}
+ x \calp _{1432} ] \prod _{A } ( { \vt [ { \e _A \atop 0} ] (\tau _A
) \over F ^\ud (x) } ) \cr && +T_{4321} [ (1-x) \calp _{1234} + x
\cals _{1432} ] \prod _{A } ( { \vt [ { \e _A \atop 0} ] (\tau _B )
\over F ^\ud (1-x) } ) \bigg ) ,\eea where \bea && \calp _{1234} =
(u_1 ^T \g ^0 u_2) (u_3 ^T \g ^0u_4) , \ \cals _{1234} = ( u_1^T \g ^0
\g ^\mu u_2) (u_3 ^T \g ^0 \g _\mu u_4) .  \eea The label $ A$ in the
products $\prod _A$ now runs over the directions of the wrapped
internal sub-torus $ T^2 _K $ common to the $ D7_I , \ D7_J$-branes,
corresponding to $ A= K_m, \ [m=1,2] $ and $ \tau _A , \ \tau _B$
retain the same definition as above except that the volume parameter
is now given by, $ \vert L_A \vert ^2 = r_K^2 $.  The Dirac spinor
scalar quartic coupling, $\calp _{1234}$, appears because the modes in
the non-diagonal open string sector $(7_I, 7_J)$ to which the fermion
pairs couple are Lorentz scalars.  Combining the contributions from
the regions of the $x$-integral near $ x=0$ and $ x=1$, yields the
low-energy representations as infinite series of $s$-channel and
$t$-channel poles located at the string compactification modes \bea &&
\cala ' _{ (3,7_I )^4 , 0} +\cala ' _{ (3,7_I )^4 , 1} \simeq C' _{Dp}
\sum _{p _A} \bigg ( T_{1234} [\cals_{1234} { \prod _A \d ^{-(p _A +
\e _A )^2 r_A ^2} \over -s + \sum _A (p _A + \e _A ) ^2 r_A ^2 } +
\calp _{1432} { \prod _A \d ^{-p _A^2 / r^2 _A } e ^{2 \pi i p _A \e
_A } / r_A \over -t + \sum _A p _A ^2 /r_A ^2 } ] \cr && + T _{4321}
[\calp _{1234} { \prod _A \d ^{-p_A^2 / r^2 _A } e ^{2 \pi i p_A \e _A
} / r_A \over -s + \sum _A p_A^2 /r_A ^2 } + \cals_{1432} { \prod _A
\d ^{-(p_A + \e _A )^2 r_A ^2} \over -t + \sum _A (p_A + \e _A ) ^2
r_A ^2 } ] \bigg ) . \eea The normalization factor is given by the
same formula as found above.

We next discuss the $ 2 \to 2 $ body processes involving the gauge
boson pair production by fermion-antifermion annihilation and the
gauge boson pair scattering processes.  The string amplitudes have a
universal form with the model dependence residing only in the gauge
structure.  In particular, identical formulas hold in single and
multiple $ Dp$-brane models.  For the localized fermions of
intersecting brane models, the string amplitude for photon pair
production by fermion-antifermion annihilation, $\cala ' _{e e \g \g }
\equiv \cala ' (e ^+ (k_1) + e ^- (k_2) + \g (k_3) + \g (k_4) )$, is
calculated from the world sheet vacuum correlator, $ < V ^{(-\ud ) }
_{-\t } V^{(-\ud ) }_{\t } V^{(-1 )}_{A_\mu } V^{(0)} _{ A _\nu } > $.
The dependence on the interbrane angles from the correlator of a
single pair of coordinate twist fields is found to exactly cancel that
coming from contracting the corresponding pair of spinor twist fields.
The rest of the calculation is standard and gives the same result as
that obtained by dimensional reduction from the $D9$-brane
amplitude~\cite{gsw2} \bea && \cala ' _{(e e \g \g )}
= G _{ Dp} [\calT_{1234} {\cals (s,t) \over st } - \calT_{1324}{\cals
(u,t) \over ut } + \calT_{1243} {\cals (s,u) \over su } ] K
(u_1,u_2,\e _3, \e _4) , \cr && [K (u_1,u_2,\e _3, \e _4) = t\ (\bar
u_1 \epslash _3 (\kslash _2 + \kslash _4) \epslash _4 u_2) + u \ (\bar
u_1 \e _4 ( \kslash _2 + \kslash _3) \epslash _3 u_2 ),\cr && \calT
_{1234} = Trace(\l _1 \l _2 \l _3 \l _4) +Trace(\l _4 \l _3 \l _2 \l
_1) ] \eea
where the factor depending on the polarization wave functions, $ K
(u_1,u_2,\e _3, \e _4) $, is (anti)symmetric under permutations of the
(fermion) boson particle labels.  Assuming, for simplicity, the three
gauge trace factors to be equal, the factorization on the massless
pole terms identifies the normalization constant to the gauge coupling
constant in the Abelian and $U(N)$ non-Abelian group cases as, $ G _{
Dp} \calT_{1234} = 2 g_ { Dp} ^2$ and $ G_{Dp} \calT _{1234} = 2 g_ {
Dp} ^2 [ \sum _a (T^a )_{12} \cdot (T^a) _{34} + {N+1\over 2 N}
(1)_{12} \cdot (1)_{34} ] $.
  
The string amplitude for the gauge boson pair scattering process,
$\cala ' _ {\g \g \g \g } \equiv \cala ' (\g (k_1) + \g (k_2) + \g
(k_3) + \g (k_4) ) $, is of same form as that obtained in the familiar
$D9$-brane case~\cite{gsw2} \bea && \cala ' _ {(\g \g \g \g )}= G '_{
Dp} [\calT _{1234} {1 \over s t } \cals (s,t) + \calT _{1324} {1 \over
ut } \cals (u,t) + \calT _ {1243} {1 \over s u } \cals (u,s) ] K _\g (
\e _1, \e _2, \e _3, \e _4) , \cr && K _\g ( \e _1, \e _2, \e _3, \e
_4) = -{1 \over 4} [st\ (\e _1 \cdot \e _3)( \e _2 \cdot \e _4) +
\text{perms} ] + {s \over 2 } \ ( (\e _1 \cdot k_4) (\e _3 \cdot k_2)
(\e _2 \cdot \e _4) + \text{perms} ) , \eea where the normalization
constant is related to the gauge coupling constant by, $ G '_{ Dp}
\calT _{1234} =G '_{ Dp} \calT_{1324} =G '_{ Dp} \calT_{1243}= 2g_ {
Dp}^2$.

\subsection{Helicity amplitudes}
\label{subsecapp2}

To enable the comparison with experimental measurements, it is useful
to express the various string amplitudes in the spin helicity basis of
the various modes.  A convenient way to proceed is by first
establishing the correspondence dictionary between the kinematics of
string amplitudes, where all particles are incoming, with that of
physical processes, and using next the familiar crossing relations
which transform particles to antiparticles and flip the sign of
momenta and helicities.  Thus, the amplitude for fermion-antifermion
pair production, $ f^+ _1 (p_1) +f^- _2 (p_2) \to f^-_3 (p_3) + f^+ _4
(p_4) $, is obtained from the string amplitude, $ \cala ( f_1 (k_1) +
f_2 (k_2) + f _3(k_3) + f_4 (k_4) ) $, by setting, $ f_1 (k_1)= f^+_1
(p_1), \ f_2 (k_2)= f^- _2 (p_2), \ f_3 (k_3) = f^- (-p_3), \ f_4
(k_4) = f^+ (-p_4) $, which involves substituting the momenta,
kinematic variables and Dirac spinors as \bea && [k_1 , \ k_2 , \ k_3
, \ k_4 ] \Longrightarrow [p_1,\ p_2, \ - p_3, \ -p_4 ] , \cr && [s =
-(k_1+k_2)^2 ,\ t = -(k_2+k_3)^2 , \ u = -(k_1+k_3)^2 ]
\Longrightarrow [s = -(p_1 + p_2)^2, \ t =- (p_1 -p_4) ^2 , \ u =-
(p_1 -p_3) ^2 ] , \cr && [u _1(k_1) , \ u_2(k_2) , \ u_3(k_3), \ u_4(
k_4) ] \Longrightarrow [ v ^\star (p_1),\ u (p_2), \ u ^\star ( p_3),\
v(p_4) ] . \eea For the choice of kinematic variables in the center of
mass frame, $\vec p_1 = -\vec p _2 =\vec p ,\ -\vec p_3 = \vec p _4
=\vec p ' ,\ [\vec p\cdot \vec p' = \cos \t ]$ the kinematic
invariants read, $ s = 4 p^2 , \ t= -s \sin ^2 {\t \over 2} ,\ u= -s
\cos ^2 {\t \over 2} .$ The helicity polarization basis for the spin
one-half Dirac fermions is described by the familiar formulas \bea &&
e = e ^-: \ u(\vec p, \l ) = p {1\choose \l } \otimes \phi _\l (\vec
p) ;\qquad \bar e= e ^+ :\ v(\vec p, \l ) = p {-\l \choose 1 } \otimes
\phi _{-\l } (\vec p) ,\cr && [\phi _L (\vec p ) = \phi _{-1 } (\vec p
) = {-\sin {\t \over 2} \choose \cos {\t \over 2} },\ \phi _R (\vec p
) = \phi _{1 } (\vec p ) = {\cos {\t \over 2} \choose \sin {\t \over
2} }.\eea The Dirac spinor matrix elements in the configuration of
helicities for the physical process, $ f ^+ f ^- \to f ^- f ^+ $ are
given by
\begin{center}
\begin{tabular}{c|c|c|c} $ \cala ( f_1 ^{(\l _1 ) } (p_1) + f_2^{(\l _2)}(p_2)
  \to f_3^{(\l _3)} (p_3) + f_4 ^{(\l _4 )} ) (p_4) $ & $ ( +--+ )$ &
$ (+-+-)$ &$ (----) $ \\ \hline $ \cals _{1234} \equiv (\bar v (p_1)
\g ^\mu u (p_2)) (\bar u (p_3) \g _\mu v (p_4) ) $ & $ -2u$ & $ -2t$
&$ 0$ \\ $ \cals _{1432} \equiv (\bar v (p_1) \g ^\mu v (p_4))( \bar u
(p_3) \g _\mu u (p_2) ) $ & $ 2u $ &$ 0 $ &$ -2s $ \\ \hline $t \cals
_{1234} - s \cals _{1432}$ & $ 2u ^2$ & $ -2t^2$ &$ 2 s ^2$ \\
\end{tabular}
\end{center}   \vskip 0.5 cm 
The results for the other helicity configurations are inferred from
the above formulas by invoking the symmetry under space parity.
The string amplitude in the three independent helicity configurations
for the physical process, $ \bar e_{\l _1} + e_{\l _2}\to \bar e_ {\l
_4} + e_{\l _3} $, are given in the single brane and the intersecting
brane cases by \bea && (\cala ^{ ' [+-+-, +--+,----] } _{ (p,p)^4} )
_{Dp} = - G_{Dp} {\cals (s,t) \over st } [2 u^2 , -2 t^2, 2 s ^2 ] +
\text{perms} , \cr && ( \cala ^{ ' [+-+-, +--+,----]} _{f^4} ) _{ISB}
= 2 \pi g_s \calT _{1234} \calv_{1234} [-2u, -2t, 0] . \eea To account
for the electroweak symmetry breaking in the charge neutral channels
one needs to substitute the massless photon pole term by the sum of $
\g + Z $ boson pole terms. The $Z$ boson exchange contribution is
obtained from that of $\g $ exchange by the substitution, $ s \to s -
m_Z ^2 $, along with the following replacements for the chirality
couplings $ LL, RR, LR, RL $: \bea && e _f ^2 \to e _f ^2 [a _L (f)a
_L^\star (f) , \ a _ R(f) a _ R ^\star (f) , \ a _L (f) a _R ^\star
(f), \ a _R (f) a _L ^\star (f) ] , \cr && [a_L(f)= {-\ud + s_W^2
\over s_W c_W }, \ a_R(f)= {s_W \over c_W }, \ s_W= \sin \t _W, c_W=
\cos \t _W ] . \label{eqgz} \eea

The differential cross section for the spin-unpolarized Bhabha
scattering process, $ \bar e + e \to \bar e + e $, obtained by adding
to the Standard Model terms the contributions from the $\cddd =6$
contact interactions, is given by \bea && < {d \s \over d \cos \t } >
= {\a ^2 \pi \over 2 s} [u^2 \vert A_{LL} \vert ^2 + u^2 \vert A_{RR}
\vert ^2 + 2 t^2 \vert A_{RL} ^s \vert ^2 + 2 s^2 \vert A_{RL} ^t
\vert ^2 ] \cr && A_{LL} = A (e ^+_ R e^- _L \to e ^+ _R e^- _L) =
A_{SM} ^{+-+-} + {2 \eta _{LL} \over \a \L ^2 }, \ [A_{SM}^{ +-+-} =
{1\over s } + {1\over t } + a_{L} ^2 (e) ({1\over s -m_Z^2 } + {1\over
t -m_Z^2 } ) ] , \cr && A_{RR} = A (e ^+ _L e^- _R \to e ^+ _L e^- _R)
= A_{SM}^{-+-+} + {2 \eta _{RR} \over \a \L ^2 }, \ [A_{SM}^{-+-+ } =
{1\over s } + {1\over t } + a_{R} ^2 (e) ({1\over s -m_Z^2 } + {1\over
t -m_Z^2 } ) ], \cr && A_{RL} ^t = A (e ^+ _L e^- _L \to e ^+ _L e^-
_L) = A_{SM}^{----} + {\eta _{RL} \over \a \L ^2 } , \ [A _{SM}
^{----} = A _{SM} ^{++++} = {1\over t } + a_L (e) a_R (e) {1\over t
-m_Z^2 } ], \cr && A_{RL} ^s = A (e ^+ _R e^- _L \to e ^+ _L e^- _R) =
A_{SM}^{+--+} + {\eta _{RL} \over \a \L ^2 } , \ [A_{SM} ^{+--+} =
A_{SM} ^{-++-} = {1\over s } + a_R (e) a_L (e) {1\over s -m_Z^2 } ] ,
\cr && A ^t _{LR} = A (e^+_R e^-_R \to e^+_R e^-_R) = A ^{++++}_{SM} +
{\eta _{LR} \over \a \L ^2 } , \cr && A ^s _{LR} = A (e^+_L e^-_R \to
e^+_R e^-_L) = A ^{-++-} _{SM} + {\eta _{LR} \over \a \L ^2 } , \eea
where $ \a = { e^2 \over 4 \pi } $ and $ {\eta _{HH'} \over \L ^2} $
denote the coefficients of the local operators previously defined in
Eq.~(\ref{eq4fernions}), and $ a_H(e)$ designate the $Z $-boson vertex
couplings defined in Eq.~(\ref{eqgz}). We use the suffix label $s, \
t$ to distinguish the $s $- and $t$-channel pole terms.

For the fermion pair scattering processes, the correspondence
dictionary between the kinematical variables in the physical process,
$\cala ( f_1(p_1) + f_2(p_2) \to f_3(p_3) + f_4(p_4) ) $, and the
string process, $\cala (f_1 (k_1) + f_2 (k_2) + f_3 (k_3) + f_4 (k_4))
$, can be written as \bea && [k_1, k_2,k_3, k_4] \Longrightarrow
[-p_3,p_1,-p_4,p_2],\ [u_1 (k_1) , \ u_2 (k_2) , \ u_3 (k_3), \ u_4
(k_4) ] \Longrightarrow [u^\star (p_3) , \ u(p_1) , \ u^\star (p_4) ,
\ u (p_2) ], \cr && [\hat s = - (k_1+k_2)^2 , \ \hat t = - (k_2+k_3)^2
, \ \hat u = - (k_1+k_3)^2 ] \Longrightarrow [ t = -(p_1 -p_4)^2, \ u=
-(p_1-p_3)^2 ,\ s=- (p_1+p_2)^2 ] . \eea For clarity, we have
distinguished the kinematic invariants of the string theory process by
adding momentarily hat symbols.  In the center of mass frame of the
physical process, $\vec p_1 = -\vec p_2 =\vec p, \ - \vec p_3 = \vec
p_4 =\vec p ' ,\ [\vec p\cdot \vec p' = \cos \t ]$ the kinematic
invariants read, $ s= 4p^2, \ t= -s \sin ^2 {\t \over 2}, \ u = - s
\cos ^2 {\t \over 2} . $ We specialize now to the physical process,
$e_H (p_1) +q_{H'} (p_2) \to e_H (p_3)+q_{H'} (p_4) $, where the Dirac
spinor matrix elements in the single $D$-brane and the intersecting
brane cases, signalled by the suffix labels $Dp$ and $ISB$, are given
by \bea && (\hat t \cals _{1234} - \hat s \cals _{1432} ) _{Dp} = [2
s^2 , \ -2 u ^2] ; \ (\cals _{1234} )_{ISB} = [ -2 \hat u,\ -2 \hat t
]= [ -2 s,\ -2 u] , \eea with the two entries inside brackets
corresponding to the equal and unequal helicity cases, $[ H=H',\ H \ne
H'] $.
The helicity basis amplitudes for the physical process can be written
for the single $D$-brane and the intersecting brane cases as \bea &&
\cala (e_H+q_{H'} \to e_H+q_{H'} )_{Dp} = G _{Dp} [\cals (u,t) {s^2
\over tu} \calT _{1324} +{\cals } (s,u) {s \over u} \calT _{1234}
+{\cals} (t,s) {s \over t} \calT _{1243} ] { 1 \choose - u^2/s^2 },
\cr && \cala (e_H+q_ {H'} \to e_H+q_ {H'} ) _{ISB} = - C [\calv
_{1324} (u,t) \calT _{1324} +\calv _{1234} (s,u) \calT _{1234} + \calv
_{1243} (s,t) \calT _{1243} ] { 2 s \choose 2 u } , \eea with the
upper and lower entries corresponding to the configurations with equal
and unequal helicities, $ [H=H' ,\ H \ne H' ] $.

The helicity amplitudes for the gauge boson pair production process, $
e ^+ (p_1) + e ^ - (p_2) \to \g (p_3) +\g (p_4) $, are given by the
formulas \bea && \cala (\bar e ^ +_R + e ^-_L \to \g _L + \g _R) =
{u\over t } \cala (\bar e ^+ _R + e ^- _L \to \g _R + \g _L) = 2 g ^2
_{Dp} \sqrt {u \over t} [{u \over s} \cals (s,t) - \cals (u,t) + {t
\over s} \cals (s,u) ]
,\eea where we have assumed the various gauge factors to be equal.
The helicity amplitudes for the gauge boson pair scattering process
are given by the explicit formulas \bea && \cala (\g _H + \g _ {H '}
\to \g _H + \g _{H '} ) = 2 g ^2 _{Dp} {s \over tu} f(s,t,u) {1\choose
{t^2 / s ^2} } ,
\cr && [f(s,t,u) = s \cals (t,u) + t \cals (u,s)+ u \cals (s,t) ] \eea
where the upper and lower entries correspond to the configurations
with equal and unequal helicities, $ [ H=H', \ H \ne H'] $.  Note the
crossing relations, $\cala (\g _{1L} + \g _{2R} \to \g _ {3L} + \g _
{4R}) \vert _{s,t,u}= \cala (\g _{1L} + \g _{4R} \to \g _ {3L} + \g _
{2R}) \vert _{t,s,u} .$



\vskip 1 cm \begin{center} \begin{figure} \epsfxsize=6 in \epsfysize=6
in \epsffile{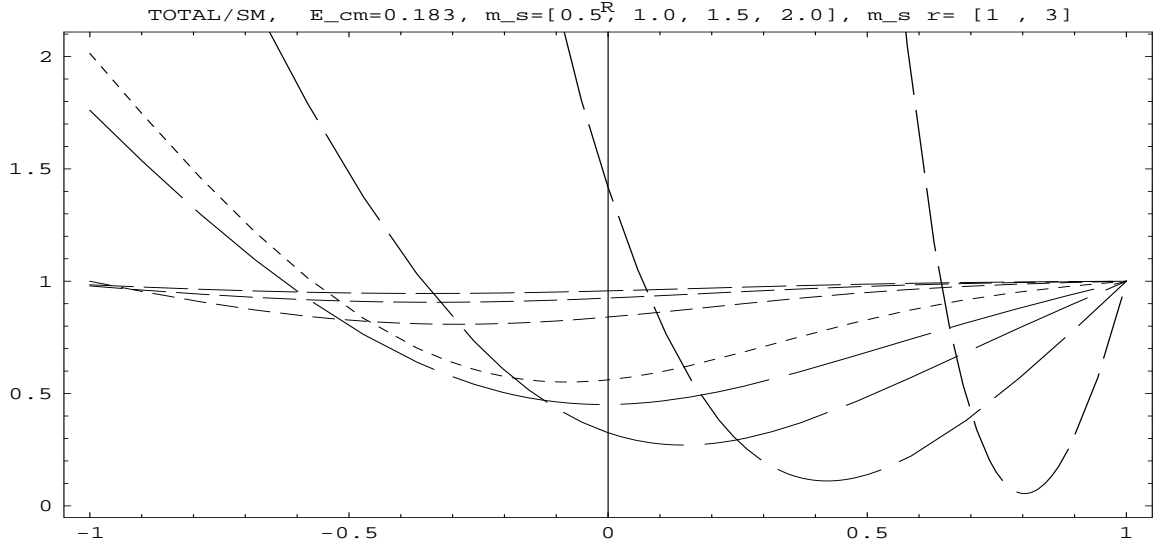}
\caption{\it The ratio $ (d \s / d \cos \t ) _{SM +contact } / (d \s /
d \cos \t ) _{SM} $ for the Bhabha scattering differential cross
section at the center of mass energy $ \sqrt s = 183 $ GeV is plotted
as a function of $ \cos \t $ using the coefficients of contact
interactions predicted in Cremades et al.,~\cite{CIM,CIM2} with the
reference set of parameters. We consider four values of the string
scale parameter, $ m_s = 0.5, \ 1.5 , \ 2., \ 3. $ TeV, and two values
of the compactification scale parameter, $ m_s r = 1 ,\ 3 .$ The group
of four lowermost curves from the bottom right corner is associated to
$ m_s r =3$, and the group of four uppermost curves close to the
horizontal axis is associated to $ m_s r =1$. The predictions for
variable string scale within each group are drawn with dashed curves
using dashings of increasing length in correspondence with the
increasing sequence of values, $m_s = 0.5, \ 1.5 , \ 2., \ 3. $ TeV.}
\label{fig4}  \end{figure} \end{center}

\vskip 1 cm \begin{center} \begin{figure} \epsfxsize=6 in \epsfysize=6
in \epsffile{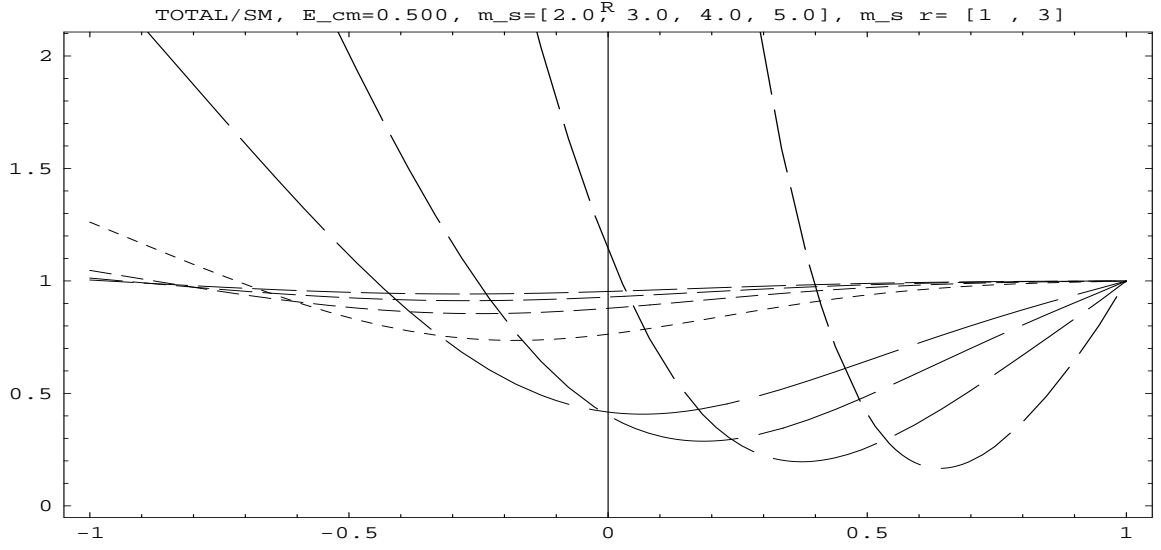}
\caption{\it The ratio $ (d \s / d \cos \t ) _{SM + contact } / (d \s
/ d \cos \t ) _{SM} $ for Bhabha scattering differential cross section
at the center of mass energy $ \sqrt s = 500 $ GeV is plotted as a
function of $ \cos \t $ using the coefficients of contact interactions
predicted in Cremades et al.,~\cite{CIM,CIM2} with the reference set
of parameters.  The group of four lowermost curves from the bottom
right corner is associated to $ m_s r =3$, and the group of four
uppermost curves close to the horizontal axis is associated to $ m_s r
=1$. The predictions for variable string scale within each group are
drawn with dashed curves using dashings of increasing length in
correspondence with the increasing sequence of values, $ m_s = 2., \
3. , \ 4., \ 5. $ TeV.}
\label{fig5}  \end{figure} \end{center}

\begin{center} \begin{figure} 
\epsfxsize= 5 in \epsfysize= 4 in \epsffile{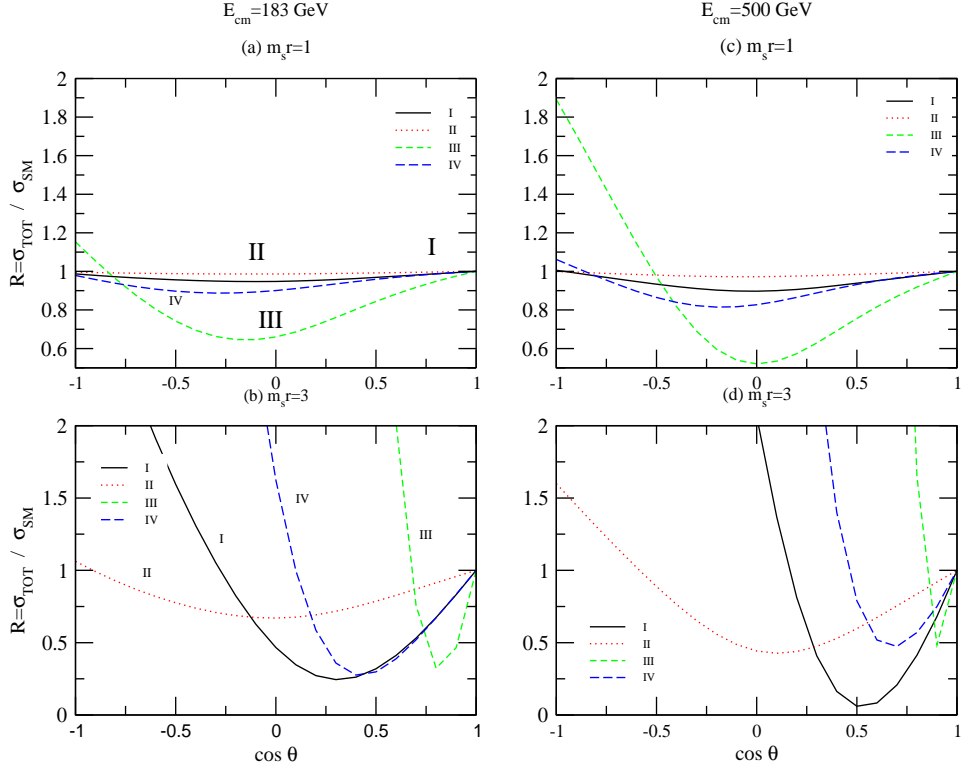}
\vskip 2 cm \caption{\it The ratio $ (d \s / d \cos \t ) _{SM +
contact } / (d \s / d \cos \t ) _{SM} $ for Bhabha scattering
differential cross section at the center of mass energies $ \sqrt s =
183.  $ GeV and $ \sqrt s = 500. $ GeV are plotted as a function of $
\cos \t $ for the string amplitudes predicted in Cremades et
al.,~\cite{CIM,CIM2} model with the reference set of parameters, using
$ g_b ^2= g_2^2 (m_Z) = 0.425$ and $\a (m_Z)= 1/127.9$.  The total
cross section is evaluated for the subtraction regularized string
amplitudes with the massless gauge boson pole terms replaced by the
corresponding pole terms at the physical masses of the neutral $\g,\
Z$ gauge bosons.  We consider the approximate estimate for the mixed
chirality amplitudes, $ G^{LR} ,\ G^{RL} $, setting these to a
constant multiple of the pure chirality amplitudes, $G^{LL}, \
G^{RR}$, defined by the parameterization, $ G^{LR} = G^{RL} = x G^{LL}
= x G^{RR}$, with the two extreme numerical values $ x= \ud , \ 5$.
We also consider the two values for the string mass scale, $m_s = 1, \
2 $ TeV and $m_s = 2, \ 4 $ TeV, in correspondence with the two center
of mass energies, $ \sqrt s = 0.183 $ TeV and $ \sqrt s = 0.500 $ TeV.
On the left hand side, the upper and lower panels $(a)$ and $(b)$
display the ratio for $ m_s r =1 $ and $ m_s r =3 $ at $ \sqrt s =
183.  $ GeV with the four curves $ I, \ II, \ III, \ IV $ (in full,
dotted, short-dashed, dashed lines) referring to the values of the
string mass scale and the proportionality factor between the mixed and
pure chirality amplitudes: $ (x= G^{LR} / G^{LL}, m_s / \text{TeV}) =
(\ud , 1), \ (\ud , 2), \ (5, 1), \ (5, 2) $ TeV.  On the right hand
side, the upper and lower panels $ (c)$ and $(d)$ display the ratio
for $ m_s r =1 $ and $ m_s r =3 $ at $ \sqrt s = 500.  $ GeV with the
four curves $ I, \ II, \ III, \ IV $ referring to the values of the
string mass scale and the proportionality factor between the mixed and
pure chirality amplitudes: $ (x= G^{LR} / G^{LL}, m_s / \text{TeV}) =
(\ud , 2), \ (\ud , 4), \ (5, 2), \ (5, 4) $.}
\label{fig6}  \end{figure} \end{center}

\begin{center} \begin{figure} 
\epsfxsize= 4 in
\epsffile{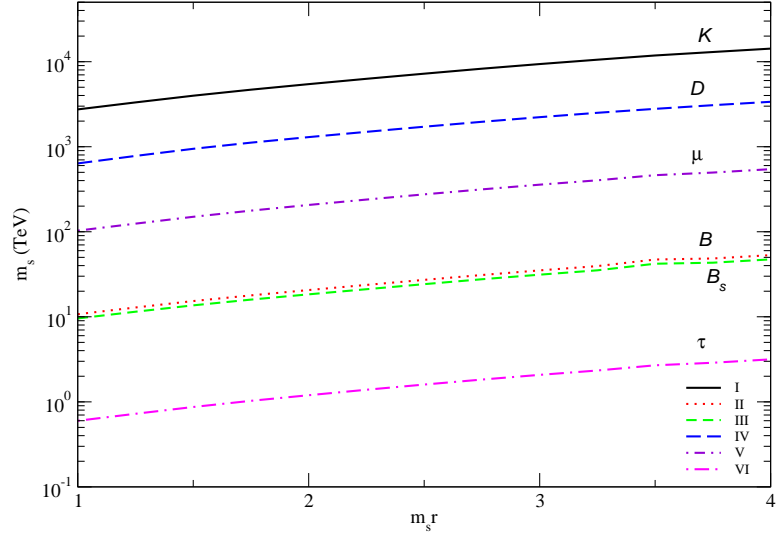}
\vskip 2 cm \caption{\it Lower bounds on $ m_s $ as a function of $
m_s r$ deduced from experimental data for the neutral mesons mass
shifts and the three-body leptonic decays of charged leptons.  The
curves labeled $ I,\ II ,\ III,\ IV $ refer to the $\vert \D F_q
\vert =2 $ mass shift observables of the neutral meson systems $
K-\bar K , B-\bar B,\ B_s-\bar B_s,\ D-\bar D$ using the experimental
inputs quoted in Eq.~(\ref{eqmesons}).
The curves labeled  $ V,\ VI $ refer to the $\vert \D F_l \vert =1 $
observables for the charged leptons three-body decay rates, $\mu \to e
+ e + \bar e ,\ \tau \to e + e + \bar e $ using the experimental
inputs quoted in Eq.~(\ref{eqleptons}).
} \label{fig7} \end{figure} \end{center}

\end{document}